\documentclass[%
 reprint,
 amsmath,amssymb,
 aps,
]{revtex4-2}

\pdfoutput=1

\usepackage{amssymb} 
\usepackage{amsthm}
\usepackage{amsmath}
\usepackage{graphicx}
\usepackage{subcaption}
\usepackage{dcolumn}
\usepackage{bm}
\usepackage{cancel}
\usepackage{float}
\usepackage{epstopdf}
\usepackage{hyperref}
\usepackage{url}
\usepackage[svgnames] {xcolor}
\usepackage{soul}

\usepackage{caption}
\usepackage{ragged2e}
\DeclareCaptionJustification{justified}{\justifying}
\hypersetup{colorlinks=true, linkcolor=blue, urlcolor=blue, citecolor=blue}

\def\chisq{\ensuremath{\chi^2}}
\def\ndof{\ensuremath{n_\text{dof}}}

\begin{document}

\title{Resolving Radiogenic Heterogeneity in the Earth's Mantle with Multi-Site Geo-neutrino Measurements}

\author{M. Leyton}
\affiliation{Division of Natural Sciences, Pasadena City College, Pasadena, CA, 91106}

\author{S. T. Dye}
\affiliation{Department of Physics and Astronomy, University of Hawaii, Honolulu, HI, 96822 USA}

\date{\today}

\begin{abstract}
\vspace{1mm}
\noindent

Abstract: {We present a global fit to geo-neutrino measurements at four underground sites (Kamioka, Gran Sasso, Sudbury, Jiangmen), combining signal rates with site-specific predictions of the lithospheric contribution to constrain the quantity and distribution of radioactivity in the Earth's mantle. Assuming a uniform distribution of radioactivity in the mantle, we find a geo-neutrino rate of $S_\text{mantle} = 14.8 \pm 4.5$~NIU when the lithospheric predictions are allowed to vary within their uncertainties. This uniform-mantle signal rate corresponds to radiogenic heating of $18.5 \pm 5.6$~TW, assuming $\text{Th/U} = 3.77$ and $\text{K/U} = 13,000$ as the ratios of the mass-fraction abundances of the main heat-producing elements. 


The uniform-mantle signal, and hence its radiogenic heating, is resolved at $\sim3.2\sigma$. However, when subtracting the lithosphere prediction from the total measured signal, there is noticeable tension between the inferred mantle signals at Jiangmen and Kamioka ($\Delta S = 24.6 \pm 13.8$~NIU, non-zero at $\Delta \chi^2=3.2$, or $1.8\sigma)$, which may indicate non-uniform geo-neutrino production in the mantle. We use $\Delta S$ as the figure of merit for the heterogeneous models examined here.



We explore six families of models to describe the heterogeneous spatial distribution of radiogenic elements in the Earth's mantle. Each model contains anomalous bodies or reservoirs carrying an excess of heat-producing elements over a depleted mantle (DM) background, with total inventory set by bulk-silicate-Earth mass balance and by a geochemical Th/U constraint. For each model, we compute the geo-response at each detector site and fit the resulting signals to the four measurements. A total of 968,274 individual model configurations are defined by varying the parameter space for each family, and then tested against the figure of merit. The results of $\sim$3 million distinct fits show that no plausible mantle or lithosphere model applied uniformly to all four sites reproduces the difference between the inferred mantle signals at the Jiangmen and Kamioka sites. Only local, site-specific prescriptions are able to reach the observed difference, suggesting that the Kamioka--Jiangmen tension is due to a local effect or statistical scatter, rather than large-scale mantle structure.

These results show that current geo-neutrino measurements begin to probe lateral heterogeneity in the mantle, and that the present tension between results at Kamioka and at Jiangmen is not explained by plausible variations of the lithosphere or by any of the large-scale mantle structures examined here, including seismic velocity anomalies and subducted slabs. Because Kamioka and Jiangmen sample largely overlapping mantle volumes, any deep, laterally extended source makes similar contributions to both sites. A contrast of the observed size thus requires additional heat-producing material concentrated much closer to one of the sites.
}
\end{abstract}

\maketitle

\section{Introduction}

Geo-neutrinos emitted by the $\beta^{-}$ decay chains of $^{238}$U and $^{232}$Th are routinely observed in large electron antineutrino detectors~\cite{abe_etal_2022, agostini_etal_2020, SNO+:2025xxx, wang_etal_2026}. The geo-neutrinos emitted by the $\beta^{-}$ decays of $^{40}$K, as well as the decay chain of $^{235}$U, however, remain undetected since their energies fall below the detection threshold. Together, U, Th and K are the main producers of radioactive heat in the Earth. Since the geo-neutrinos emitted by their decays carry information on the elemental abundances and distributions within the Earth, they provide important clues about the composition of the Earth's interior and the size and sources of its internal heating due to radioactivity, both in the current epoch and throughout its evolution.

Radiogenic heat production is one of the principal energy sources of the Earth’s interior, yet its magnitude remains poorly constrained. The Earth’s total surface heat flow is $47 \pm 2$~TW~\cite{Davies2010}, but heat-flow measurements are unable to distinguish radiogenic heat from primordial heat leftover heat from Earth's formation, heat flowing from the core into the mantle~\cite{buffett_2003}, or secular cooling of the mantle. Estimates of the radiogenic share therefore come from compositional models of the bulk silicate Earth (BSE), which predict between $\sim$10 and 33~TW~\cite{sramek2013}. Of this share, the lithosphere contributes $7.9^{+1.8}_{-1.3}$~TW~\cite{huang_etal_2013}, leaving the mantle contribution uncertain by more than a factor of ten, from about 2 to 25~TW. The mantle share governs the secular cooling rate of the mantle, the vigor of mantle convection and plate tectonics, and the heat flux across the core–mantle boundary available to sustain the geodynamo. While rock samples typically reach only the crust and the shallow mantle, geo-neutrino measurements can directly constrain the present-day U and Th inventory of the whole Earth, including the deep mantle.

In this paper we combine existing geo-neutrino measurements in order to constrain the quantity and distribution of radioactivity in the Earth’s mantle. After summarizing the current measurements and lithosphere predictions in Section~\ref{sec:geonu-observations}, we fit a uniform mantle to the data in Section~\ref{sec:homogeneous_fit_withpulls} and test its stability. In Section~\ref{sec:naive_mantle_model}, we calculate the inferred mantle signals at each observation site and introduce a figure of merit for testing heterogeneous models. After defining six families of heterogeneous models in Sections \ref{sec:llvp_models}--\ref{sec:local_models}, we describe the common likelihood framework and plausibility criteria in Sections~\ref{sec:likelihood_framework} and \ref{sec:scan_ranges}. We present the results in Section~\ref{sec:results} and discuss their statistical context in~\ref{sec:limitations}. Our conclusions are summarized in Section~\ref{sec:conclusions}.


\section{Geo-neutrino observations\label{sec:geonu-observations}}

Assessments of the geo-neutrino rate now exist from underground observatories in Kamioka, Japan~\cite{abe_etal_2022}, Gran Sasso, Italy~\cite{agostini_etal_2020}, Sudbury, Canada~\cite{SNO+:2025xxx} and Jiangmen, China~\cite{wang_etal_2026}. The observatories at these four sites sample the geo-neutrino rate from the continents of Asia, Europe, and North America, spanning a wide range of longitudes at latitudes between $\sim20^\circ$ N and $\sim50^\circ$ N. They detect electron antineutrinos through inverse beta decay (IBD) on free protons, measuring the energy, but not the direction, of electron antineutrinos above the $1.8$-MeV interaction threshold. As such, they are sensitive to several unstable daughters in the decay series of $^{238}$U and $^{232}$Th, allowing assessments of the geo-neutrino rate from those heat-producing elements. Assessments rely on detailed knowledge of the magnitudes and energy spectra of the subtracted background sources and on a fixed mixture of contributions from $^{238}$U and $^{232}$Th to define the shape of the geo-neutrino energy spectrum and to reduce the uncertainties of the reported signal rate. 

Table~\ref{tab:geonu_rates_sites} presents the measured total geo-neutrino rates and the predicted lithosphere rates at the four observation sites. Using geophysical and geochemical models, together with dedicated field studies, the geo-neutrino rates from the lithosphere at the four observatories are predicted with a precision of typically $\pm5$ Neutrino Interaction Units (NIU)~\footnote{1 NIU is equal to 1 IBD interaction per year exposure of $10^{32}$ free protons and is equivalent to $1$~Terrestrial Neutrino Unit (TNU)}. Since the four sites are embedded in different crustal terranes (e.g.\ Sudbury sits on the Canadian Shield, while Jiangmen lies among extensive U- and Th-rich granitic intrusions~\cite{Gao2020JULOC}), the lithosphere beneath each site has its own U and Th content and Th/U ratio. Separating the mantle signal therefore depends on modelling each site’s lithospheric contribution accurately. The mantle contribution, by contrast, has a simple expectation: if the mantle is laterally uniform, every site should record nearly the same mantle signal. We test this hypothesis first by fitting a single mantle rate to all four sites. 

\begin{table}[H]
\caption{Rates of geo-neutrinos from uranium and thorium, $R$(U+Th) in NIU, at the four observation sites. The total rates are assessed at Kamioka, Japan by KamLAND~\cite{abe_etal_2022}, at Gran Sasso, Italy by Borexino~\cite{agostini_etal_2020}, at Sudbury, Canada by SNO+~\cite{SNO+:2025xxx}, and at Jiangmen, China by JUNO~\cite{wang_etal_2026}. The lithosphere rates are predicted at Kamioka, Japan~\cite{huang_etal_2013}, at Gran Sasso, Italy~\cite{agostini_etal_2020}, at Sudbury, Canada~\cite{strati_etal_2017,huang_etal_2013,laursen_etal_2019}, and at Jiangmen, China~\cite{Gao2020JULOC} using geophysical and geochemical models, together with dedicated field studies. The lithosphere prediction at Kamioka described elsewhere~\cite{abe_etal_2022} is $25.1 \pm 5.3$ NIU, exceeding our selected value by $2.4$ NIU, even without including a contribution from the Continental Lithospheric Mantle (CLM).}
\setlength{\tabcolsep}{4pt}
\begin{tabular}{l l l l l}
\hline\noalign{\smallskip}
$R$(U+Th)              & Kamioka                & Gran Sasso              & Sudbury                & Jiangmen  \\
\hline\noalign{\smallskip} 
Total                & $30.7^{+4.5}_{-4.5}$   & $47.0^{+8.6}_{-8.1}$    & $48^{+14}_{-12}$       & $73^{+11}_{-11}$ \\ [3pt]
Lithosphere          & $22.7^{+4.9}_{-4.1}$   & $25.9^{+4.9}_{-4.1}$    & $34.2^{+9.2}_{-5.3}$   & $40.4^{+5.6}_{-5.0}$ \\ [3pt]
\hline\noalign{\smallskip}
\end{tabular}
\label{tab:geonu_rates_sites}
\end{table}

\section{Uniform Mantle Fit\label{sec:homogeneous_fit_withpulls}}

The simplest test of the data is whether a single, uniform mantle signal common to all four sites, added to the site-specific lithosphere predictions, can account for the four observed rates. If the distribution of heat-producing elements in the mantle is uniform, then the mantle signal rate is essentially constant at near-surface sites, and any variations in the total measured rates are due only to differences in the lithosphere rates and to statistical fluctuations. Since the lithospheric Th/U differs between sites, we fit the combined U+Th rates rather than the individual U and Th rates, and we allow the lithosphere prediction at each site to vary within its quoted uncertainty. 

We fit a single parameter ($S_\text{mantle}$) against 4 data points ($R$(U+Th) per site), with 4 nuisance parameters ($\delta_{i}$) corresponding to a lithosphere normalization pull, in units of $\sigma$, at each site. These nuisance parameters allow the predicted lithosphere contribution at each site to shift up or down from its nominal value, as presented in Table~\ref{tab:geonu_rates_sites}. By applying these values per site, we prevent mistakenly attributing a site-specific crust effect to the mantle. The $\chi^2$ has 8 terms (4 data + 4 nuisance penalties) and 5 parameters, leaving $\ndof = 3$. The best-fit rate, $\chisq/\ndof$ and $p$-value are the reference quantities used to compare subsequent fits in this study.

\begin{figure}
\centering
\includegraphics[trim = 0mm 0mm 0mm -12mm, clip, scale=0.43]{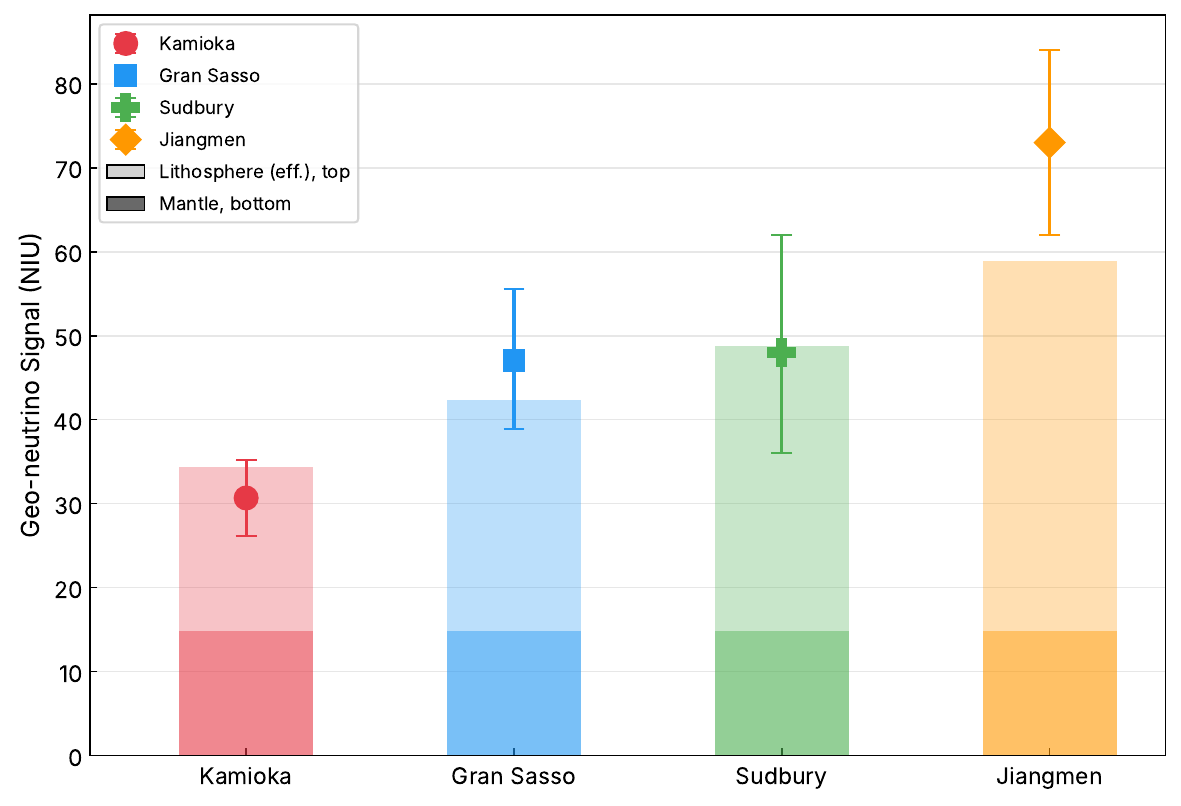}
\caption{Data vs.\ best-fit predictions per site, assuming a uniform mantle, and split into effective lithosphere and mantle contributions. The lithosphere predictions at each site have been adjusted by the corresponding best-fit nuisance parameter. The best-fit mantle signal at all four sites is $S_\text{mantle} = 14.8 \pm 4.5$~NIU, corresponding to an equivalent heat of $18.5 \pm 5.6$~TW, assuming $\text{Th/U} = 3.77$ and $\text{K/U} = 13,000$ as the ratios of the mass-fraction abundances.}
\label{fig:homo_mantle_results}
\end{figure}

\begin{figure}
\centering
\includegraphics[trim = 0mm 0mm 0mm 0mm, clip, scale=0.47]{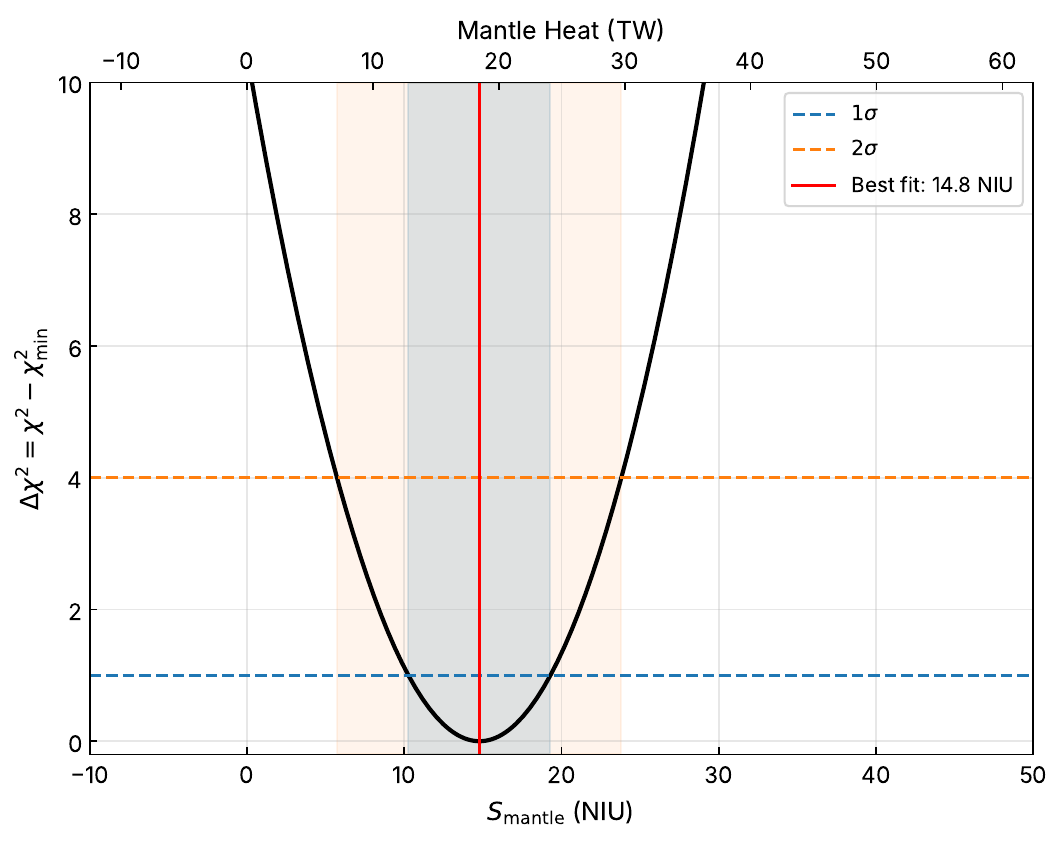}
\caption{Profile-likelihood fit of the mantle signal rate at the four sites, assuming a uniform mantle. The best-fit rate is $S_\text{mantle} = 14.8$~NIU, while the $1\sigma$ and $2\sigma$ crossing points are [10.3, 19.3]~NIU and [5.7, 23.8]~NIU, respectively. The equivalent mantle heat (TW) is shown on the upper axis.}
\label{fig:homo_mantle_PL}
\end{figure}

The best-fit uniform mantle signal using Minuit (Migrad/Hesse) is $S_\text{mantle} = 14.8 \pm 4.5$~NIU, with \chisq/\ndof\ = 3.78/3 and $p=0.286$. The lithosphere nuisance pulls are $-0.75 \sigma$, $+0.35 \sigma$, $-0.03 \sigma$, and $+0.66 \sigma$ at Kamioka, Gran Sasso, Sudbury, and Jiangmen, respectively. These results are shown in Figure~\ref{fig:homo_mantle_results}. A profile-likelihood analysis yields a $\pm 1 \sigma$ interval of [10.3, 19.3]~NIU, and a $\pm 2 \sigma$ interval of [5.7, 23.8]~NIU, consistent with the results using Minuit. Figure~\ref{fig:homo_mantle_PL} shows the results of the profile-likelihood fit assuming a uniform mantle. 

\begin{figure}
\centering
\includegraphics[trim = 0mm 0mm 0mm 0mm, clip, scale=0.45]{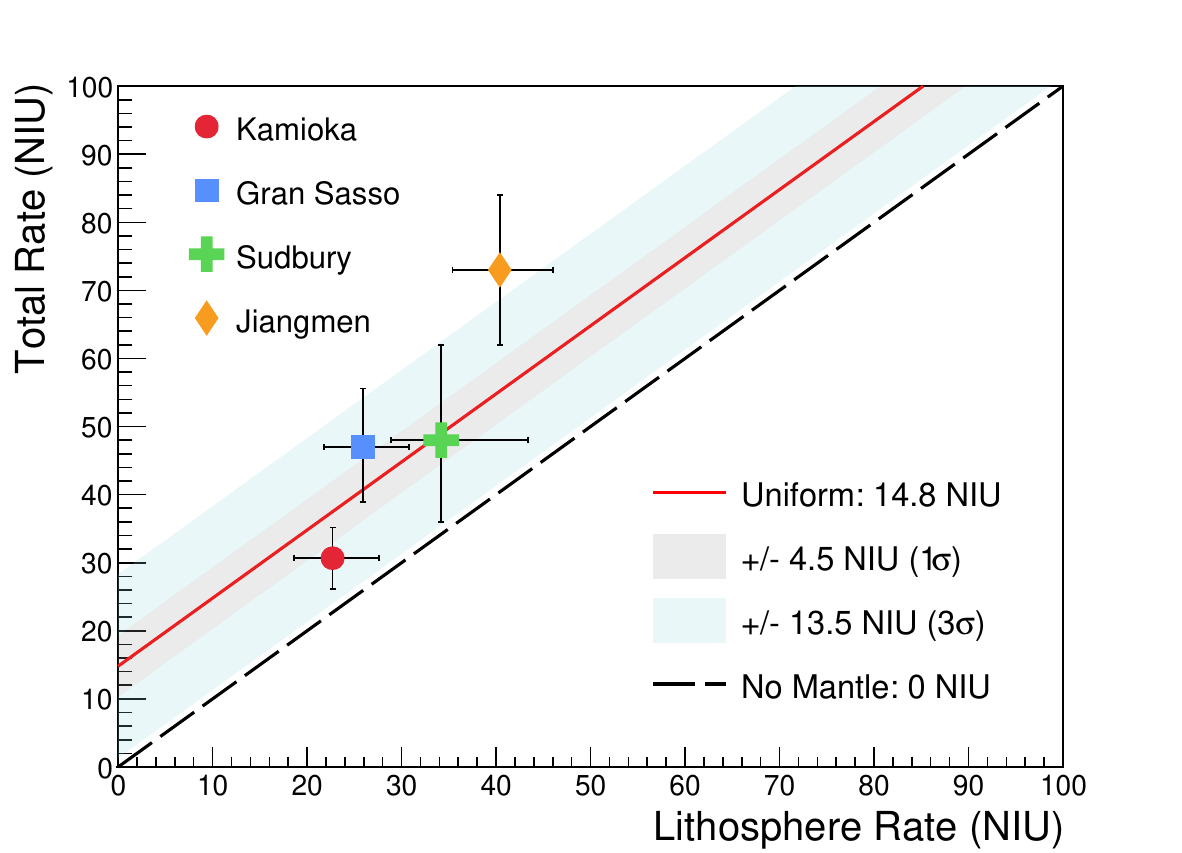}
\caption{The assessed geo-neutrino rates are plotted versus the predicted geo-neutrino rates from the lithosphere at the four observation sites, together with the best-fit mantle signal $S_\text{mantle} = 14.8 \pm 4.5$~NIU (red line). The parallel line passing through the origin represents a mantle rate of 0~NIU. The constant positive offset estimates the geo-neutrino rate from the uniform mantle with a significance of $3.2 \sigma$.}
\label{fig:mantle_resolution}
\end{figure}

Figure~\ref{fig:mantle_resolution} plots the assessed geo-neutrino rates versus the predicted lithosphere rates at the four underground observation sites, together with the best-fit mantle signal. If U and Th are uniformly distributed in the mantle, these multi-site data resolve the near-surface mantle rate with significance $3.2 \sigma$ ($\sqrt{\Delta \chi^2}$ at $S_\text{mantle} = 0$). Assuming $\text{Th/U}=3.77$ and $\text{K/U}=13,000$, the best-fit mantle rate corresponds to $18.5 \pm 5.6$~TW of radiogenic heating in the mantle. Adding $7.9^{+1.8}_{-1.3}$~TW of radiogenic heating in the lithosphere~\cite{huang_etal_2013} to that in the uniform mantle totals $26 \pm 6$~TW.

To explore the influence of any single-site measurement on these results, we repeat the fit with all possible combinations of measurements from three out of four sites. Table~\ref{tab:uniform_mantle_fits} summarizes the results of these ``leave-one-out" fits. In each fit, the uniform mantle rate is resolved at $2.4\sigma$ or better. Moreover, the uniform mantle rates of all fits are consistent within uncertainties. Removing Kamioka or Jiangmen sharply reduces $\chi^2$ (from 3.78 to 1.02 or 1.38, respectively), while removing Gran Sasso or Sudbury barely changes it. Without Kamioka, the mantle signal rises to 23~NIU, indicating that the other three sites prefer a higher mantle signal, but Kamioka alone pulls the common value down. 

Equivalently, the leave-one-out fits can be read as giving one site its own free mantle signal offset. The largest improvement is at Kamioka, where the data prefer an offset of $-15.1 \pm 9.2$~NIU ($\Delta \chi^2 = 2.76$ relative to the uniform mantle, 1.7$\sigma$), where a negative offset means the site requires less mantle signal than the common value. These results suggest a mild ($\approx 1.7\sigma$) preference for a site offset at Kamioka or Jiangmen and that both sites are worth investigating, in terms of lithosphere systematic uncertainties, detector-specific effects, or a genuine geophysical signal not captured by the current uniform mantle model.

\begin{table}[H]
\caption{Summary of uniform mantle fits, with lithosphere predictions at each site allowed to vary. Fit results are presented including measurements from all four sites and for all possible combinations of measurements from three out of four sites.}
\setlength{\tabcolsep}{3pt}
\begin{tabular}{l c c c c c}
\hline\noalign{\smallskip}
& $S_\text{mantle}$ (NIU) & \chisq & \ndof\ & Prob. & Sig. \\
\hline\noalign{\smallskip} 
All four sites	& $14.8 \pm 4.5$ & 3.78	& 3	& 0.29 & $3.29\sigma$\\
w/o Kamioka	& $23.0 \pm 6.7$	& 1.02	& 2	& 0.60 & $3.4\sigma$\\
w/o Gran Sasso	& $12.9 \pm 5.1$ & 3.21	& 2	& 0.20 & $2.53\sigma$\\
w/o Sudbury	& $14.9 \pm 4.7$	& 3.78	& 2	& 0.15 & $3.17\sigma$\\
w/o Jiangmen & $12.0 \pm 4.9$	& 1.38	& 2	& 0.50 & $2.45\sigma$ \\
\hline\noalign{\smallskip}
\end{tabular}
\label{tab:uniform_mantle_fits}
\end{table}


\section{Naive Mantle Model\label{sec:naive_mantle_model}}

The simplest view of the data is to calculate the ``naive" mantle signal at each site, defined as the observed geo-neutrino rate minus the predicted lithosphere signal, $S_\text{NM} = S_\text{obs} - S_\text{LS}$. It is the mantle signal ``left over" after subtracting each site's lithosphere prediction. We also calculate the combined statistical-plus-lithosphere uncertainty and the ``allowed" mantle range by sweeping the lithosphere prediction across its quoted range with the observed rate held at its central value.



\begin{figure} [H]
\centering
\includegraphics[trim = 0mm 0mm 0mm 0mm, clip, scale=0.4]{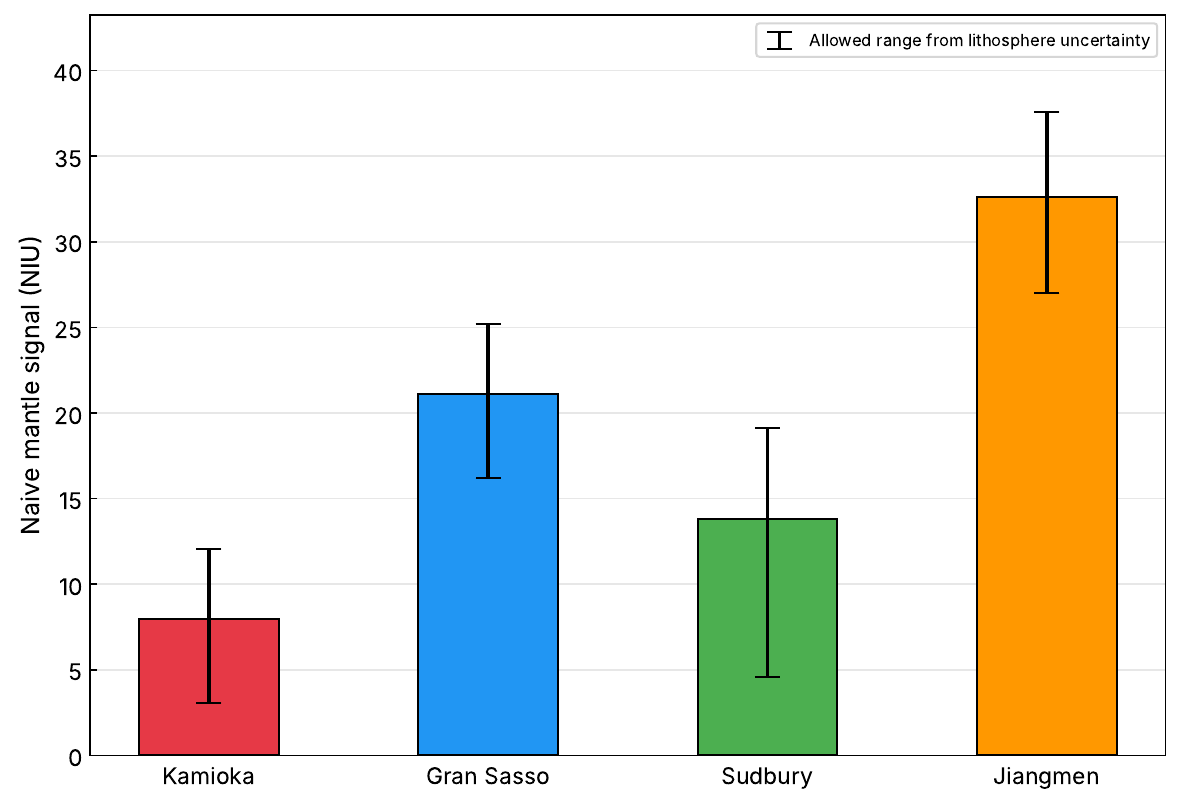} \\
\caption{Mantle signal (NIU) in the naive model after site-specific lithosphere subtraction. Error bars show the lithosphere-prediction uncertainty only; statistical errors are not included.} 
\label{fig:naive_mantle}
\end{figure}

The results are shown in Table~\ref{tab:naive_mantle} and Figure~\ref{fig:naive_mantle}. The naive values ignore the common mantle assumption and the fit's ability to shift the lithosphere prediction at each site within its range (the nuisance pulls). Sites with a large lithosphere error relative to the naive mantle signal cannot independently say much about the mantle; the lithosphere model, not the measurement, then limits the inference.

\begin{table*}[t]
\caption{Naive mantle signal $S_\text{NM}$ at each site, defined as the observed geo-neutrino rate minus the predicted lithosphere signal, $S_\text{obs} - S_\text{LS}$.}
\setlength{\tabcolsep}{8pt}
\begin{tabular}{l c c c c c c c}
\hline\noalign{\smallskip}
Site & $S_\text{obs}$ & $S_\text{LS}$  & $S_\text{NM}$ & stat.\ err. & LS err. &   combined & allowed signal range (NIU)  \\
 & & & & & & LS + stat.\ err. & (LS uncertainty only) \\
\hline\noalign{\smallskip}
Kamioka & 30.7     & 22.7   & 8.0       & 4.5      & 4.5      &  6.4   & [    3.1,    12.1 ] \\
Gran Sasso & 47.0   & 25.9 & 21.1       & 8.3      & 4.5       & 9.5  & [   16.2,    25.2 ] \\
Sudbury  & 48.0     & 34.2 & 13.8      & 13.0      & 7.2      & 14.9   & [    4.6,    19.1 ] \\
Jiangmen &  73.0     & 40.4  & 32.6      & 11.0      & 5.3      & 12.2   & [   27.0,    37.6 ] \\
\hline\noalign{\smallskip}
\end{tabular}
\label{tab:naive_mantle}
\end{table*}


At Kamioka, $S_\text{NM} = 8.0$~NIU, with a range of 3.1--12.1 NIU from the uncertainty of the predicted lithosphere rate alone. At Jiangmen, $S_\text{NM} = 32.6$~NIU (27.0--37.6 NIU), while the inferred mantle rates at Gran Sasso and Sudbury fall between. With statistical and lithosphere uncertainties combined, the naive Jiangmen $-$ Kamioka difference is $\Delta S_\text{NM} = 24.6 \pm 13.8$~NIU, approximately 1.8$\sigma$ from 0. Physically, a difference of this size between these two sites is difficult to produce. The mantle signal at a detector is an integral over thousands of kilometers of mantle, weighted by distance as $1/L^{2}$. A laterally uniform (spherically symmetric) mantle therefore gives essentially the same signal everywhere. Kamioka and Jiangmen are separated by $25.8^\circ$ (about 2,900 km), and both in East Asia, so they sample largely overlapping mantle volumes. Deep, large-scale structures such as LLVPs, subducted slabs or other velocity anomalies are therefore seen in nearly the same way at both sites.

This makes the difference 
\begin{eqnarray}
\label{eqn:FOM}
    \Delta S = S_\text{mantle}(\text{Jiangmen}) -S_\text{mantle}(\text{Kamioka})
\end{eqnarray}
a natural figure of merit for assessing and comparing heterogeneous mantle models. $\Delta S$ does not depend on the total abundance of U and Th in the mantle, only on how these elements are arranged relative to the two sites. The overall mantle abundance, the neutrino oscillation probability, and the conversion to NIU change the signals at both sites by nearly the same proportion, meaning that they largely drop out from the calculation. A model that matches every site’s absolute rate within errors but predicts $\Delta S \approx 0$ has not accounted for the main feature of the data. Conversely, a model can only reproduce $\Delta S$ by putting heat-producing material much closer to Jiangmen than to Kamioka. $\Delta S$ therefore turns the comparison of heterogeneous models into a direct question: does a given distribution of radiogenic elements create a site-to-site contrast of the observed size, and at what geochemical or geophysical cost?


\section{Heterogeneous Mantle Models\label{sec:heterogeneous_mantle_model}}

We now fit the four-site dataset to an extensive suite of physically motivated models describing various heterogeneous spatial distributions of radiogenic elements in the Earth's mantle (or lithosphere). Each model defines a set of anomalous bodies within the mantle (or lithosphere) that carry an excess of heat-producing elements over a depleted-mantle (DM) background. Models are sorted into six families, according to how the anomalous bodies are defined:
\begin{itemize}
    \item \textit{LLVP models} (\textbf{H}): two large low-velocity provinces (LLVPs) defined according to existing literature (Section~\ref{sec:llvp_models}).
    \item \textit{Slow-velocity anomalies} (\textbf{S}): regions drawn from tomographic vote maps, either attached to the core-mantle boundary (CMB) or chosen by global depth windows (Section~\ref{sec:sf_anomalies}).
    \item \textit{Fast-velocity anomalies} (\textbf{F}): regions drawn from tomographic vote maps, with and without the reference LLVPs (Section~\ref{sec:sf_anomalies}).
    \item \textit{Subducted slabs} (\textbf{B}): data from the Atlas of the Underground~\cite{vanderMeer2018} are used to define anomalous bodies in the mantle with typical slab composition (Section~\ref{sec:subducted_slabs}).
    \item \textit{Lithosphere components} (\textbf{L}): each layer of the crust and lithospheric mantle scaled within its uncertainties (see Section~\ref{sec:lithosphere_components}).
    \item \textit{Local site-specific models} (\textbf{C}): anomalies selected only within a given radius of a detector, plus a free mantle offset at Jiangmen (Section~\ref{sec:local_models}).
\end{itemize}
In every family of models, the mantle consists of a set of bodies $b$ (LLVPs, velocity anomalies, slabs, etc.) embedded in a depleted mantle (DM), which is defined as the remainder of the mantle. More information on each of these families can be found in their respective sections. 

For each model configuration, we compute the geo-response $G_{b,k}$ (in g/cm$^{2}$) of each body $b$ at a detector site $k$, as
\begin{equation}
    G_{b,k} = \int_b \frac{\rho \, dV}{4 \pi L^2},
    \label{eqn:geo-response}
\end{equation}
where the integral is over the volume of the body, $\rho$ is the density of a volume element $dV$, and $L$ is its distance from the detector. The total geo-response at a detector site $k$ is then given by

\begin{equation}
    G_{\text{tot},\,k} = \sum_{i} G_{i,k} = G_{\text{DM},\,k} + \sum_{b} G_{b,k},
\end{equation}
using
\begin{equation}
    M_{\text{tot}} = \sum_{i} M_{i} = M_{\text{DM}} + \sum_{b} M_{b},
\end{equation}
where the sum $i$ is over all bodies + DM, $M_{\text{tot}} = 3.978 \times 10^{24}$~kg is the total mass of the mantle, independent of body shape and composition, and $M_\text{DM}$ is the mass of the DM. The masses of all bodies + DM sum to a fixed total since the DM takes up the mass released when the anomalous bodies shrink

The total inventories of U and Th are set by bulk-silicate-Earth (BSE) mass balance and by a geochemical Th/U constraint, described in Section~\ref{sec:geochemical_constraint}. Writing the elemental abundance for each body $b$ and element $X$ as
\begin{equation}
    a_b(X) = a_\text{DM}(X) + x_b(X),
\end{equation}
where $a_\text{DM}(X)$ is the abundance of element $X$ in the DM and $x_b(X)$ is its excess over DM, then the predicted mantle signal at a site $k$ is the abundance-weighted sum over all bodies + DM:
\begin{equation}
    S_{\text{mantle, }k} = \langle P_{ee} \rangle \sum_{X=\text{U,Th}} F_X \sum_{i} G_{i,k}\,a_{i}(X),
\end{equation}
where $\langle P_{ee} \rangle =0.548$ is the average electron-antineutrino survival probability and $F_{X}$ is the NIU-per-unit-abundance factor of isotope chain $X$. Abundances $a_i(X)$ for each of the mantle reservoirs (bodies or DM) and for the six families of models studied here are computed from one or more of the materials listed in Table~\ref{tab:abundances}.

\begin{table}
\caption{U and Th abundances used to compute the average abundance $a_i(X)$ of a given mantle reservoir (anomalous body or DM), dependent on its composition. OCR and BCC values are from \cite{huang_etal_2013}; GLOSS values are from \cite{Plank1998}. Range of DM abundances is from \cite{workman_hart_2005,salters_stracke_2004,Arevalo2010Chemical,huang_etal_2013} and includes uncertainties.}
\setlength{\tabcolsep}{3pt}
\begin{tabular}{l c c}
\hline\noalign{\smallskip}
Material & U (ppb) & Th (ppb) \\
\hline\noalign{\smallskip}
Recycled oceanic crust (OCR) & 70 & 210 \\
Bulk continental crust (BCC) & 1310 & 5610 \\
Subducting sediment (GLOSS) & 1680 & 6910 \\ 
Depleted mantle (DM) & 3.2--12 & Th/U = 2.5--2.9 \\
\hline\noalign{\smallskip}
\end{tabular}
\label{tab:abundances}
\end{table}

Differences in density between anomalous bodies and the DM are not modeled here. All mantle reservoirs use the laterally uniform density from PREM~\cite{dziew81}. Intrinsic density contrasts of a few percent would change both body masses and geo-responses proportionally and negligibly. 

The resulting mantle signals at each site, $S_{\text{mantle,}\,k}$, are then fit to the four measurements, with profiled lithospheric nuisance parameters. Across families, we scan over both ``plausible" and ``non-plausible" parameter space (defined in Section~\ref{sec:scan_ranges}), DM abundances, vote-map threshold, and body composition and geometry. We then compare the result of each fit against the figure of merit defined in Equation~\ref{eqn:FOM}. 

\section{Vote Count and Blend Ratio Parameterization}

To characterize the size, shape, and composition of the anomalous bodies in the first three families of models (\textbf{H}, \textbf{S}, \textbf{F}), as well as the last family of models (\textbf{C}), we employ a geometric (Vote Count) and compositional (Blend Ratio) parameterization, as described below. This parameterization is more suitable for studying the range of tomographic models employed by these families, compared to the usual seismic velocity perturbation parameter ($\delta V_s/V_s$). 

\begin{itemize}
\item Vote Count (``VC", integer range 1--18): a discrete tomographic-consensus parameter that counts how many independent global shear-velocity models agree that a given region belongs to an anomalous body. Low VC includes any region where even a minority of models see a signal (large volume, lower confidence). High VC keeps only the ``core" of the anomaly where nearly all models agree (small volume, high confidence). In other words, a higher VC corresponds to a stricter consensus and a smaller, but more confident, anomalous region. VC therefore controls the geometry and volume of each anomalous body and sets how strongly the predicted mantle signal at each site responds to a unit change in composition.
 
\item Blend Ratio (``BR", 0--100\%): the compositional mixing fraction, in percent, between background DM and a fully enriched endpoint composition in the anomalous bodies. Setting $\text{BR}=0$ leaves all anomalous bodies at their base composition, while 100\% assumes the anomalous bodies have fully attained their enriched endpoint composition, typically that of bulk continental crust (BCC). The exact definition of BR depends on the family of models; more details can be found in the respective section for each family. Note that the BR parameter is itself constrained by bulk silicate Earth (BSE) geochemistry (Section~\ref{sec:geochemical_constraint}). 

\end{itemize}

VC and BR control different aspects of the model. VC changes only the geometric response, i.e.\ how much a given anomalous body can contribute to the signal at each site. BR is the single compositional parameter of the model and changes only the elemental abundance used within the anomalous body defined by VC. This factorization allows us to derive an independent, purely geochemical BR constraint (as a function of VC) from BSE mass balance, entirely decoupled from the detector data. 

\section{Geochemical VC-BR Constraint\label{sec:geochemical_constraint}}

Although the total amounts of Th and U are not well constrained, the bulk silicate Earth's Th/U mass ratio is a well-measured quantity, $3.77^{+0.12}_{-0.07}$~\cite{wipperfurth_etal_2018}, essentially independent of the detector data studied here. This places a constraint on which (VC, BR) combinations are geochemically self-consistent, supplying an independent handle from bulk silicate Earth (BSE) geochemistry. This is especially helpful since the detector data alone may be unable to uniquely pin down both VC and BR, given that large, weakly-enriched anomalous bodies and small, strongly-enriched anomalous bodies look very similar to the four detector sites.

Here we use the measured Th/U mass ratio to derive and parameterize the relationship between VC and BR using a Padé [2/2] rational function (quadratic-over-quadratic) with 5 parameters:

\begin{equation}
    \text{BR(VC)} = \frac{p_0 + p_1 \cdot \text{VC} + p_2 \cdot \text{VC}^2}{1 + q_1 \cdot \text{VC} + q_2 \cdot \text{VC}^2}.
\end{equation}

Three contours are calculated corresponding to Th/U = 3.77 (nominal), as well as symmeterized $\pm 1 \sigma$ bounds at 3.67 and 3.87. The constraint itself is derived from a BSE mass-balance argument, with abundances of the anomalous bodies set to those described in each family's respective section, typically that of BCC (see Section~\ref{sec:reservoir_composition}). The BSE Th/U mass ratio is calculated as the total mass of Th divided by the total mass of U, summed over the lithosphere and every mantle reservoir (anomalous bodies + DM):

\begin{equation}
    \left( \frac{\text{Th}}{\text{U}} \right)_\text{BSE} = \frac{M_\text{Th,LS}+ \sum_{i} M_{\text{Th, }i}}{M_\text{U,LS}+ \sum_{i} M_{\text{U, }i}}.
\end{equation}
The element mass $M_{X,\,i}$ of each mantle reservoir $i$ is its total mass times an elemental abundance (mass fraction) for a given reservoir: $M_{X,\,i}=M_i \, a_i(X)$, where $M_i$ are the reservoir masses, calculated as described in Section~\ref{sec:heterogeneous_mantle_model} and interpolated between the tabulated integer VC values. The abundances $a_i(X)$ are calculated from the admixture of each reservoir, as a function of BR, for each isotope $X=\{\text{U,\,Th}\}$. 
Since the crust and DM terms do not depend on BR, and the anomalous body terms are linear in BR, both the total Th mass and total U mass are exactly linear functions of BR. Pinning their ratio to the target BSE value is therefore a single linear equation in BR.

The geochemical constraint is folded into the fit as a soft Gaussian penalty, rather than a hard cut. The detector data and the geochemical prior are explicitly weighed against each other via the constraint weight $w$ and the constraint is then added to the total $\chi^2$ as:

\begin{equation}
    \chi^2_\text{geo} = w \cdot \left( \frac{\text{BR} - \text{BR}^\text{nom}_\text{pred}(\text{VC})}{\sigma_\text{band}(\text{VC})} \right) ^2,
\end{equation}
where $\text{BR}^\text{nom}_\text{pred}(\text{VC})$ is the nominal curve prediction at a given VC, $\sigma_\text{band}(\text{VC})$ is the half-width of the $\pm 1\sigma$ band at that VC value, and $w$ is an adjustable constraint weight set equal to 1, so that the constraint is treated as a proper $1\sigma$ Gaussian pull. 

\begin{figure} [H]
\centering
\includegraphics[trim = 0mm 0mm -12mm 0mm, clip, scale=0.39]{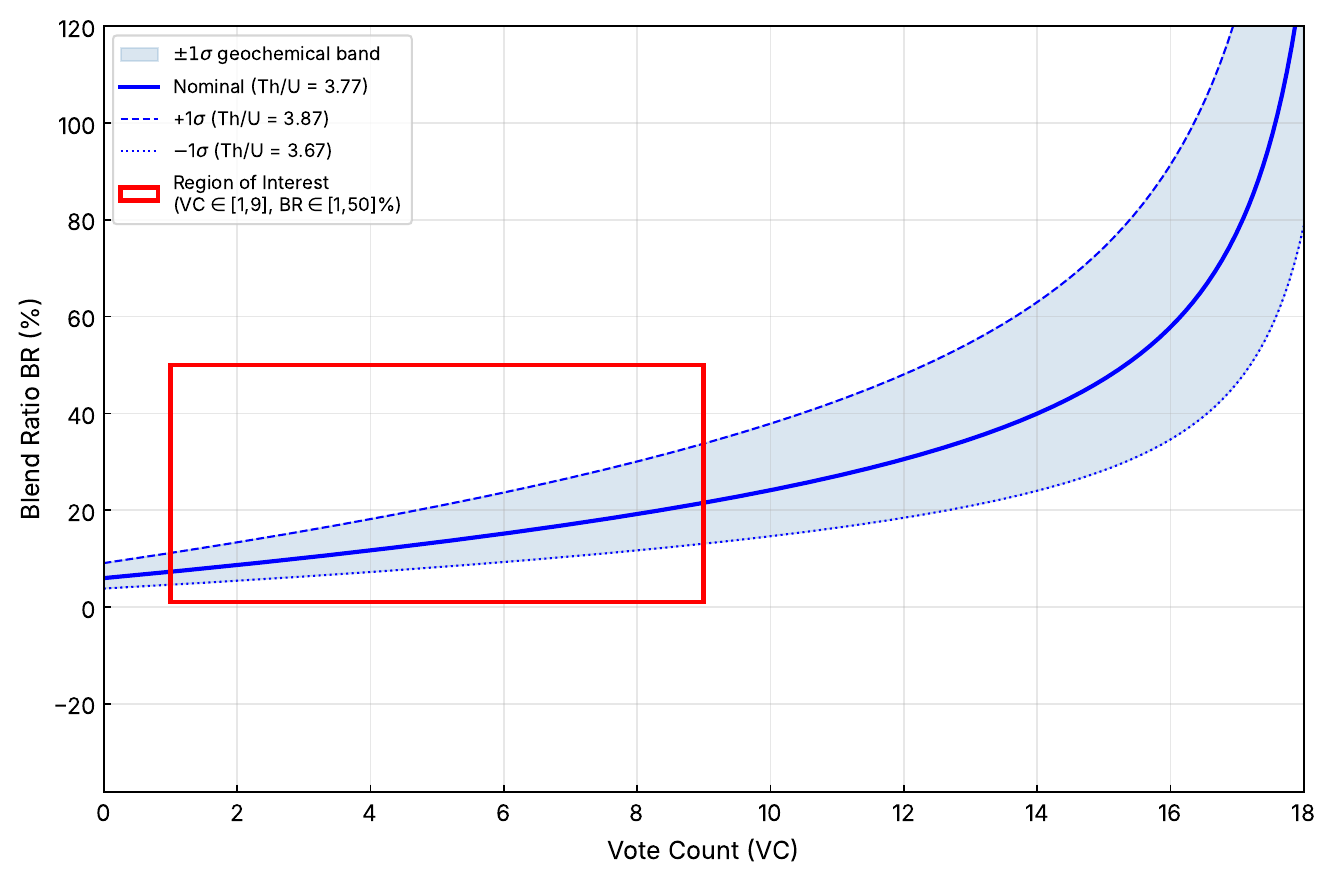}
\caption{An example BR(VC) geochemical constraint curve for Th/U = 3.77 (nominal), as well as symmeterized $\pm 1 \sigma$ bounds at 3.67 and 3.87. The red box marks the region of interest for this analysis.}
\label{fig:geochem_constraint}
\end{figure}

Figure~\ref{fig:geochem_constraint} shows an example BR(VC) constraint curve (nominal and $\pm 1 \sigma$ band) over its complete fitting domain (VC=1--18). The red box marks the region of interest for this analysis for visual context. Note that BR is constrained in each family to prevent a negative DM fraction within the anomalous bodies. Also, a high BR can signify an implausible quantity of continental crust in the mantle. The full curve is shown here only for completeness and to ensure a smooth function beyond the edges of the fit window. 

We now describe each of the family of models in more detail, starting with the LLVP models.

\section{LLVP Models (H)\label{sec:llvp_models}}

Seismic tomography of the lowermost mantle consistently resolves two deep mantle structures called Large Low-Shear-Velocity Provinces (LLVPs). These broad, seismically anomalous piles sit on the core-mantle boundary (CMB), one beneath Africa and the other in the Central Pacific. Their reduced shear-wave velocity is generally attributed to some combination of higher temperature and different composition, possibly enriched in iron and in incompatible elements, relative to the surrounding depleted mantle (DM). Because U and Th are strongly incompatible elements, a compositionally enriched LLVP is expected to carry a disproportionate share of the mantle's radiogenic heat budget. These LLVPs provide some motivation for modeling the mantle not as a spatially-averaged, homogeneous entity, but as a heterogeneous one with lateral variations that mimic the resolved LLVP structures.

In the LLVPs models (\textbf{H}) studied here, the mantle is partitioned into three distinct reservoirs: the depleted mantle (DM), which forms the bulk of the convecting mantle, and the African and Pacific LLVPs.

\subsection{Reservoir Shape and Tomographic VC Prior\label{sec:VCprior}}

An analysis of seismic models explores the maximum heights of the Pacific and African LLVPs, finding $\sim700$--800~km and $\sim1600$--1800~km, respectively~\cite{yuan_li_2022}. Earlier results find similar heights, with the Pacific LLVP extending $740$~km~\cite{he_wen_2009} and the African LLVP $1300$~km~\cite{wang_wen_2007} above the CMB. Our reference LLVP model uses 720 and 1600~km for the heights of the Pacific and African LLVPs, respectively, but we scan over a range of height configurations, including sets with both LLVPs 200, 400, 600, and 800~km tall.

\begin{figure*}
    \centering
    \begin{subfigure}[t]{0.47\textwidth}
        \centering
        \includegraphics[trim = 0mm 130mm 0mm 0mm, width=0.99\linewidth]{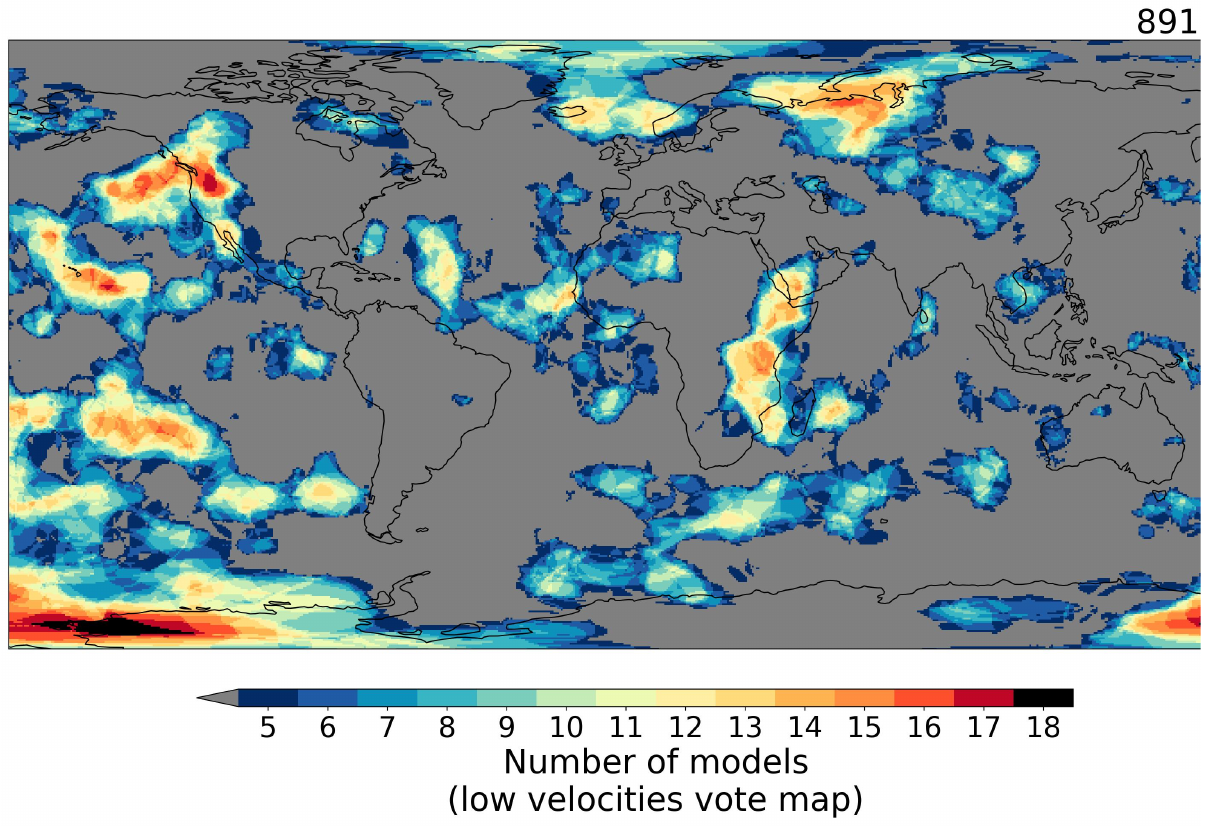} 
        \caption{$2000$ km above CMB} \label{fig:blobs1}
    \end{subfigure}
    \begin{subfigure}[t]{0.47\textwidth}
        \centering
        \includegraphics[trim = 0mm 130mm 0mm 0mm, width=0.99\linewidth]{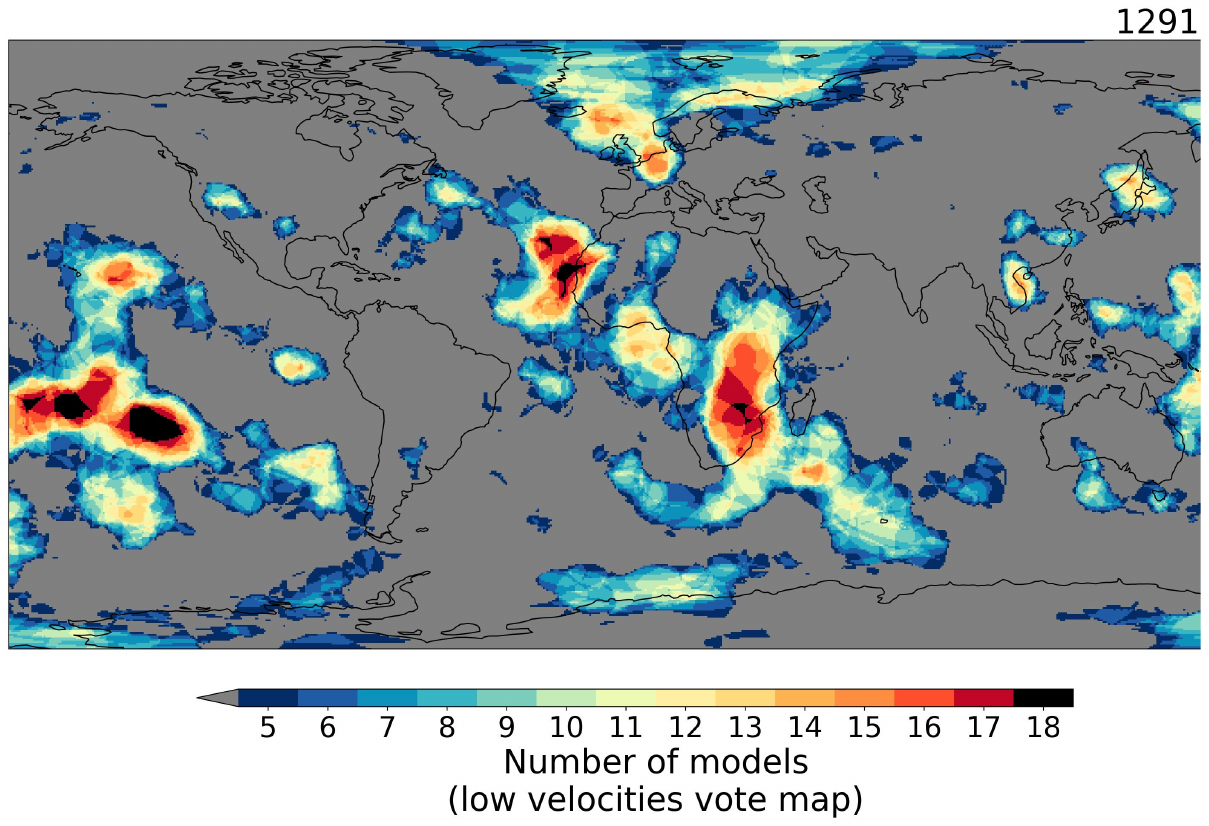} 
        \caption{$1600$ km above CMB} \label{fig:blobs2}
    \end{subfigure}
    \begin{subfigure}[t]{0.47\textwidth}
        \centering
        \includegraphics[trim = 0mm 130mm 0mm 0mm, width=0.99\linewidth]{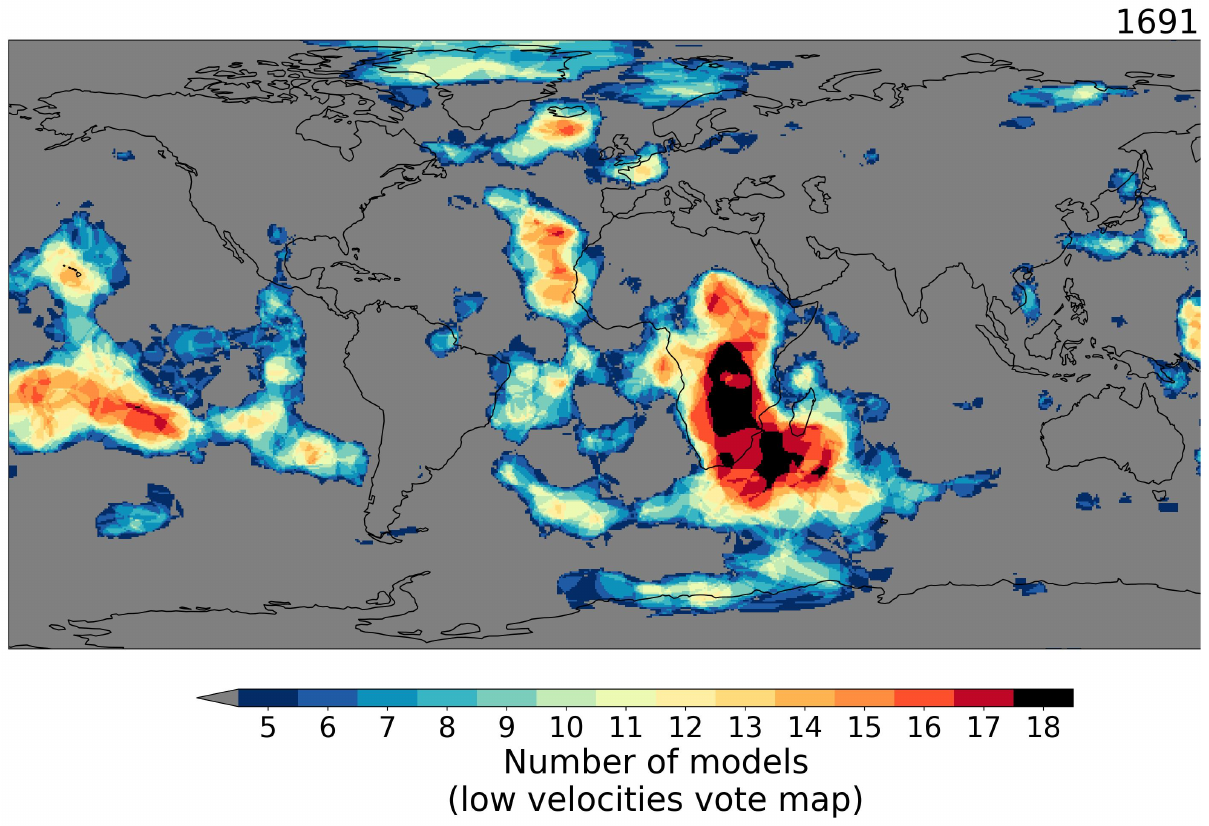} 
        \caption{$1200$ km above CMB} \label{fig:blobs3}
    \end{subfigure}
    \begin{subfigure}[t]{0.47\textwidth}
        \centering
        \includegraphics[trim = 0mm 130mm 0mm 0mm, width=0.99\linewidth]{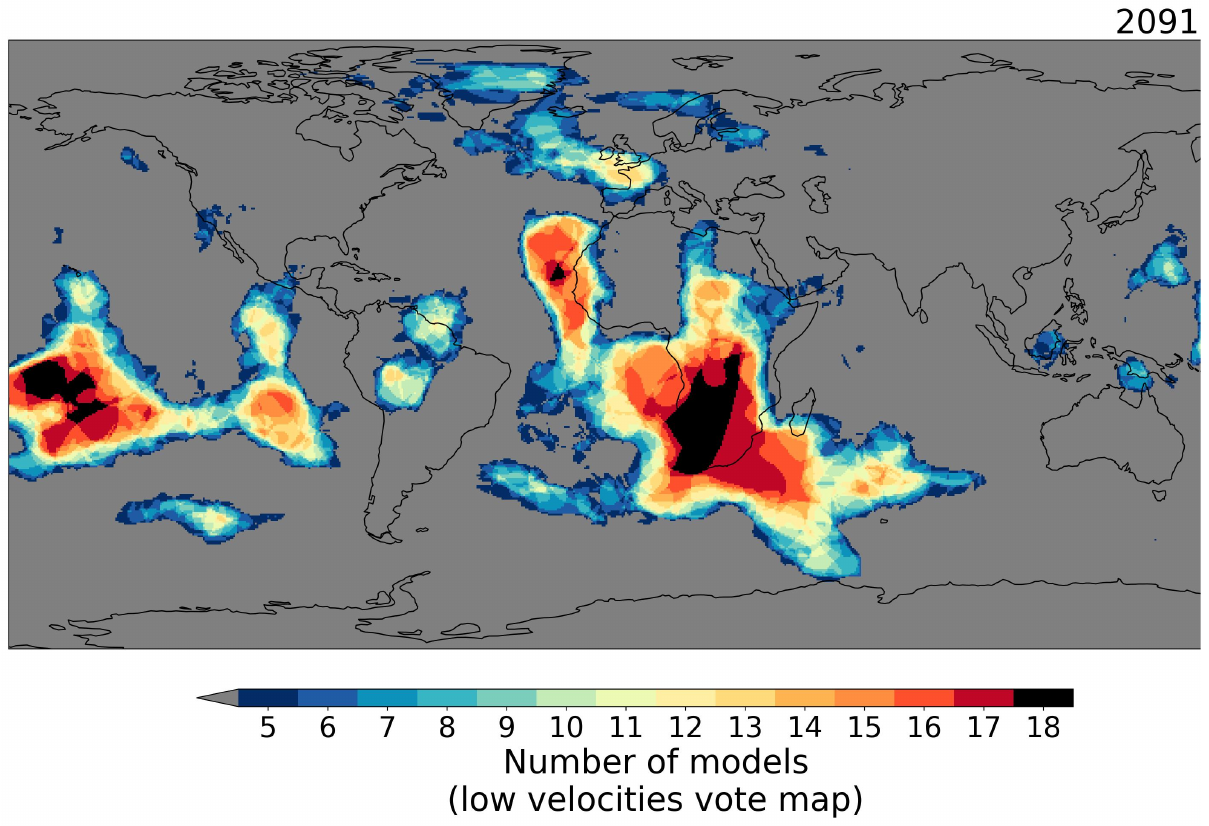} 
        \caption{$800$ km above CMB} \label{fig:blobs4}
    \end{subfigure}
    \begin{subfigure}[t]{0.47\textwidth}
        \centering
        \includegraphics[trim = 0mm 130mm 0mm 0mm, width=0.99\linewidth]{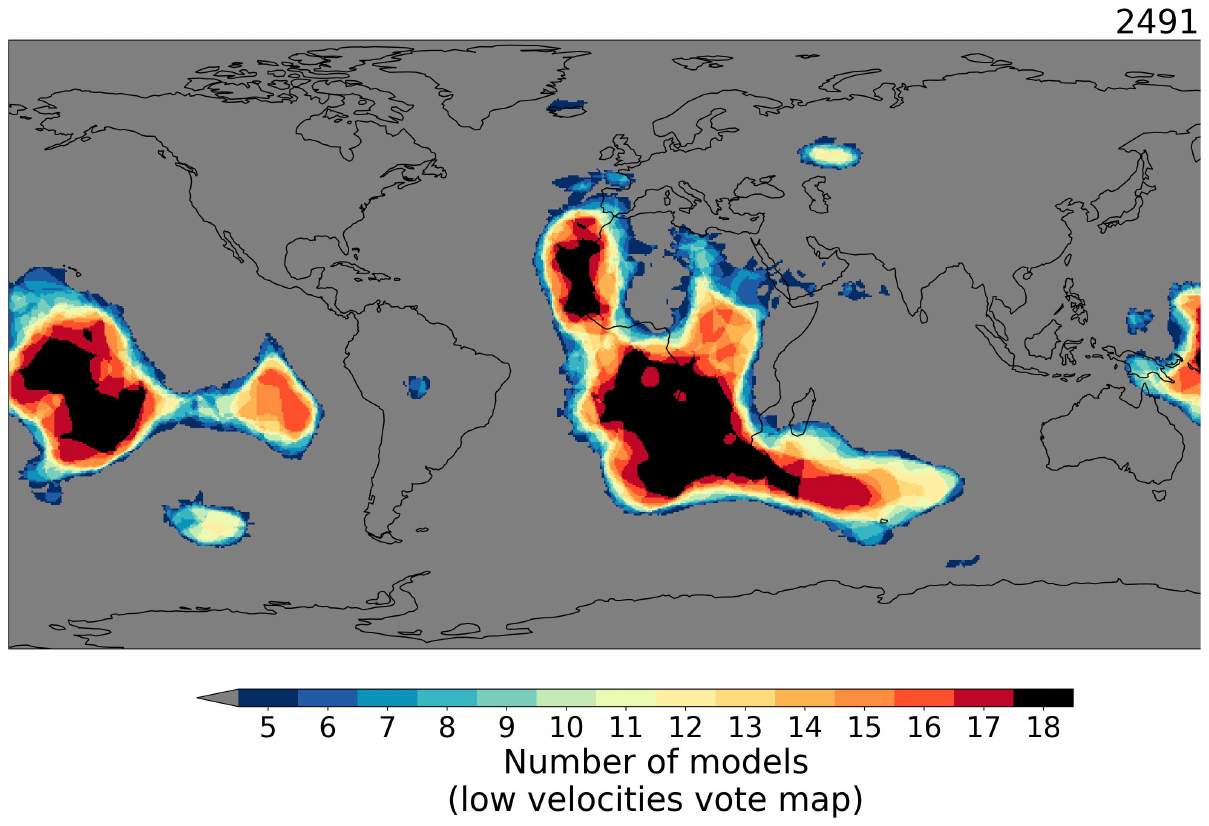} 
        \caption{$400$ km above CMB} \label{fig:blobs5}
    \end{subfigure}
    \begin{subfigure}[t]{0.47\textwidth}
        \centering
        \includegraphics[trim = 0mm 130mm 0mm 0mm, width=0.99\linewidth]{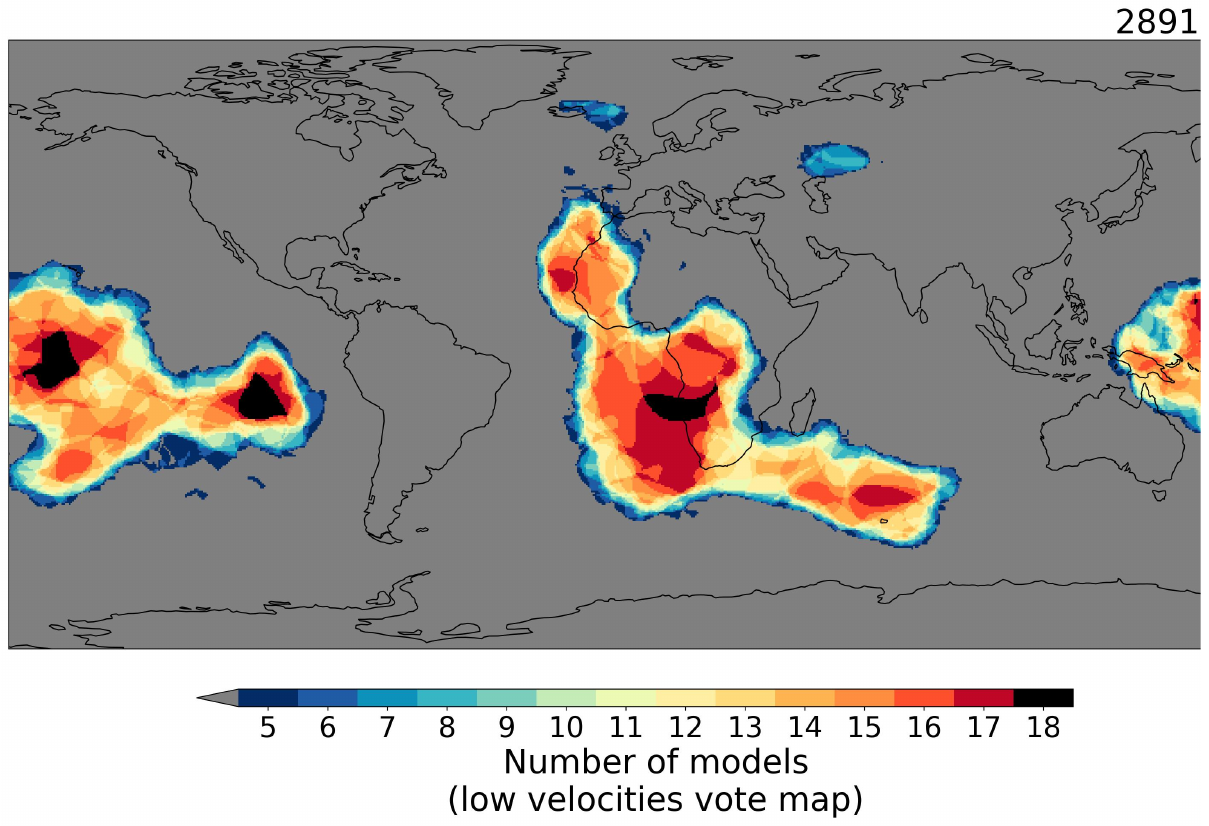} 
        \caption{$0$ km above CMB} \label{fig:blobs6}
    \end{subfigure}
    \caption{Tomographic vote maps with Vote Count (VC) $\geq$ 5 from 18 independent global shear-velocity models, compiled on SubMachine~\cite{SubMachine} with data from \cite{Shephard_2017}, at various depths representative of our reference model for the Pacific and African LLVPs.} 
    \label{fig:llvp_vote_maps}
\end{figure*}

Tomographic vote maps used to define the shape of the LLVPs are taken from 18 independent global shear-velocity models compiled on SubMachine~\cite{SubMachine} with data from \cite{Shephard_2017}. Figure~\ref{fig:llvp_vote_maps} shows a series of vote maps at various depth slices for $\text{VC} \geq 5$.

While the geochemical Th/U mass constraint (see Section~\ref{sec:geochemical_constraint}) is helpful for tying BR to VC, it is still necessary to anchor VC. Without this anchor, VC will walk to whichever boundary the optimizer prefers, with an error bar spanning nearly the whole scan window. 
Cottaar \& Lekic~\cite{Cottaar2016} introduce a ``vote count" tomographic-consensus quantity when comparing 5 independent global shear-velocity models. For every location in the lowermost mantle, they count how many of the 5 models assigned it to the LLVP cluster and report that number as the vote count. Their study is also the standard reference for the vote-map concept used by Shephard et al.~\cite{Shephard_2017}, who compiled vote maps from a larger (14 in their original study) set of tomographic models. Although Shephard et al.\ caution that their vote maps were not designed to characterize the LLVPs, we use the vote maps only to define how the LLVP geometry changes with the agreement threshold, and we fix the overall LLVP size independently by anchoring VC to the Cottaar \& Lekic LLVP mass fraction via the prior described here. As shown in Section~\ref{sec:results}, the figure of merit is insensitive to the exact choice of vote threshold.

\begin{figure}
\centering
\includegraphics[trim = 0mm 0mm -8mm 0mm, clip, scale=0.42]{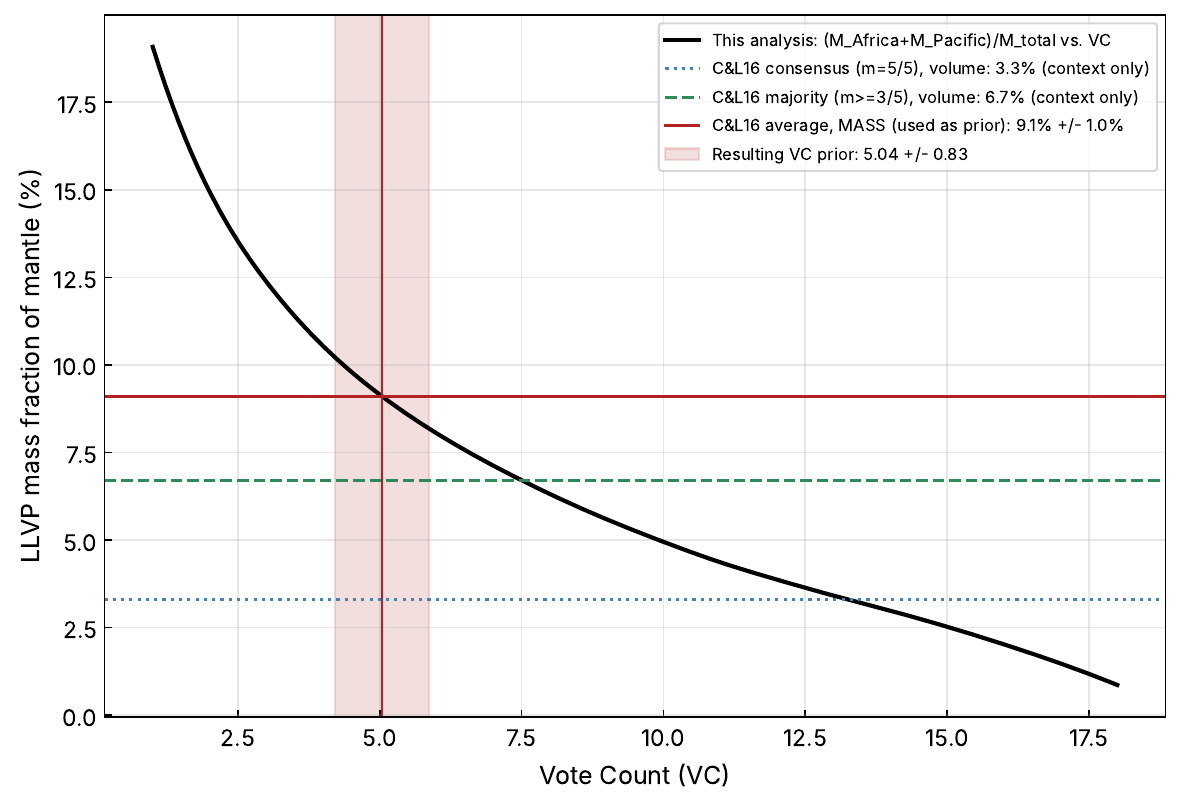}
\caption{LLVP mass fraction of mantle (\%) as a function of Vote Count (VC), used to translate the mass fraction measured by Cottaar \& Lekic~\cite{Cottaar2016} into a VC prior. This calculation results in VC = $5.04 \pm 0.83$ for our reference LLVP model.}
\label{fig:VCprior}
\end{figure}

Cottaar \& Lekic report the total LLVP fraction of the mantle under three vote thresholds, giving both a volume fraction ($8.0\% \pm  0.9\%$) and a mass fraction ($9.1\% \pm 1.0\%$) for their headline average estimate. For a direct comparison between models, we use Cottaar \& Lekic's mass fraction since no unit conversion or density assumption is needed. However, their VC result cannot be used directly since the model presented here uses a different (finer) discretization than theirs. Instead we translate their physical estimate of the LLVP mass fraction onto our VC axis using our mass tables to relate VC to physical LLVP extent. We solve for the VC that results in an LLVP mass fraction of 0.091 (see Figure~\ref{fig:VCprior}) and propagate the $\pm 1\%$ mass-fraction uncertainty through the local slope of that same curve to calculate the uncertainty. We calculate the VC prior for our reference LLVP model to be $5.04 \pm 0.83$.

As additional confirmation, Panton et al.~\cite{panton_etal} use $\text{VC}\geq5$ to define the boundary of the LLVPs. To better understand the dependence of the fit results on choice of VC, we additionally scan over all integer values of VC from 1--18.

These values are used as a Gaussian prior and added to the likelihood function anywhere the heterogeneous fits couple VC and BR. The prior here does almost all of the work of pulling VC to a well-defined interior value. The data alone barely move it, meaning that they are unable to constrain VC. However, the soft prior treatment is retained since it correctly propagates the tomographic uncertainty into BR (inflating its error by 3.5\%), not just the point estimate.

\subsection{Reservoir Composition\label{sec:reservoir_composition}}

There are several choices for the set of abundances of heat-producing elements (HPEs) in the DM. One publication describes the calculation of the abundances, which are provided in a table~\cite{salters_stracke_2004}. Another publication references the HPE abundances for the depleted Mid-Ocean Ridge Basalt (MORB), which are then reduced by a factor of $10$ to arrive at the DM abundances in~\cite{huang_etal_2013}. In our reference LLVP model, we assume that the DM surrounding the LLVPs has abundances of U and Th that are slightly higher than those in the depleted MORB mantle (DMM), $a_\text{DM}(\text{U}) = 10$~ppb, $a_\text{DM}(\text{Th}) = 27$~ppb. However, we scan over a range of DM abundances, as presented in Table~\ref{tab:abundances}. 

Studies suggest that LLVPs may partially consist of subducted slabs of oceanic crust~\cite{panton_etal,Krymarys2025}. Because the levels of U and Th in the oceanic crust (OCR) are enriched by a factor of $\sim$7--8 relative to the depleted mantle, we use a mass-weighted mixture to describe the composition of each of the reservoirs. The mixture for each reservoir is composed of three end members: depleted mantle (DM), recycled oceanic crust (OCR), and bulk continental crust (BCC), with abundances of each member given in Table~\ref{tab:abundances}. 

Matching of ultra-high-pressure shear wave velocity measurements with subducted-MORB models shows that the seismic anomaly typically observed in LLVPs can be explained by 23–33\% (by volume) oceanic crust along a cold slab geotherm, without the need for extreme thermal anomalies~\cite{Krymarys2025}. We take the midpoint of this range at 28\% as the base composition for each of the LLVPs. Although the literature value of 28\% is a volume fraction, it is used here as the mass fraction in the abundance average, neglecting the difference in densities between the OCR and DM (see Section~\ref{sec:heterogeneous_mantle_model}). 

For consistency with geochemical analyses of ocean island basalts (OIBs)~\cite{jackson_etal_2021}, the LLVPs are additionally infused with a component of bulk continental crust (BCC). The literature is genuinely split on this matter: Doucet et al.\,\cite{doucet2020} find a continental-recycling isotopic signature specific to the African LLVP, due to enrichment by subducted continental material during the assembly/breakup of Pangaea, which is absent in the Pacific domain. Other work describes it as ``older, well-mixed" material with possibly lower geologically recent enrichment in subducted oceanic crust than the Pacific LLVP~\cite{panton_etal}. In this study, the African LLVP receives the mass fraction BR (in percent) of BCC, displacing an equal mass fraction of DM. BR is left as a free parameter in our model and is the only compositional fit parameter. 
The Pacific LLVP receives a fraction $k$ of that amount. The abundances in each of the LLVP reservoirs in our reference model are thus computed as:
\begin{align}
\label{eqn:reservoir_abundances}
    a_\text{Afr} = & \,\,0.28\,a_\text{OCR} + \text{BR} \cdot a_\text{BCC}+ (0.72-\text{BR})\,a_\text{DM} \nonumber \\
    a_\text{Pac} = & \,\,0.28\,a_\text{OCR} + k \cdot \text{BR} \cdot a_\text{BCC}\, + \nonumber \\
    & \,\, (0.72-k \cdot \text{BR})\,a_\text{DM},
\end{align}
where $k=1/10$. Since both LLVPs also carry a fixed 28\% OCR admixture, a value of $\text{BR}=0$ leaves both LLVPs at their base composition (28\% OCR and 72\% DM). These same compositions enter the BSE mass balance that defines the geochemical constraint in Section~\ref{sec:geochemical_constraint}, which sums over the crust, DM, and both LLVPs. 

In our mantle model, the amount of ocean crust in the mantle grows with the mass of the LLVPs. When expressed as a fraction of the mantle mass and VC, it varies from $\sim5 \%$ at $\text{VC}=1$ to $\sim0.2$\% at $\text{VC} =18$. For comparison, the amount of ocean crust generated throughout Earth history is estimated to be 6--7\%, assuming production at a constant rate over 4.55~Ga~\cite{workman_hart_2005}. Although mineral physics studies show that the subducted oceanic crust becomes more dense than the surrounding lower mantle~\cite{hirose_etal_2005}, allowing it to naturally sink to the CMB, mantle convection calculations show that the subducted oceanic crust is preferentially entrained in mantle plumes rather than accumulated above the CMB, or in the LLVPs~\cite{li_mcnamara_2013}. 

To quantify and better understand the dependence of the fit results on these compositional choices, we also vary the LLVP OCR fraction $ \in \{ 0,\,0.23,\,0.28,\,0.33 \}$ and allow $k \in \{ 0,\,1/10,\,1/5,\,1/3,\,1/2,\,1 \}$, in addition to the DM abundances given in Table~\ref{tab:abundances}. A total of 46,656 model configurations are tested. The results presented in Section~\ref{sec:results} show that the figure of merit (Equation~\ref{eqn:FOM}) is insensitive to choice of DM abundances, OCR fraction, and $k$.

\subsection{Geo-response\label{sec:geo-response}}

Masses $M_i\text{(VC)}$ of each reservoir $i$ and corresponding geo-responses $G_{i,k}\text{(VC)}$ at each of the detector sites $k$ are calculated as described in Section~\ref{sec:heterogeneous_mantle_model}, as a function of the tomographic vote count (VC) in integer values from 1--18. A cubic spline is used to interpolate between the tabulated integer VC values so that the geometric response can be evaluated continuously across the VC range during fitting and profiling. Note, however, that Minuit treats VC as continuous, as needed for smooth gradients. The reader should round to the nearest integer for a strictly discrete physical interpretation of any reported best-fit value.

\begin{figure} [H]
\centering
\includegraphics[trim = 0mm 0mm -12mm 0mm, clip, scale=0.38]{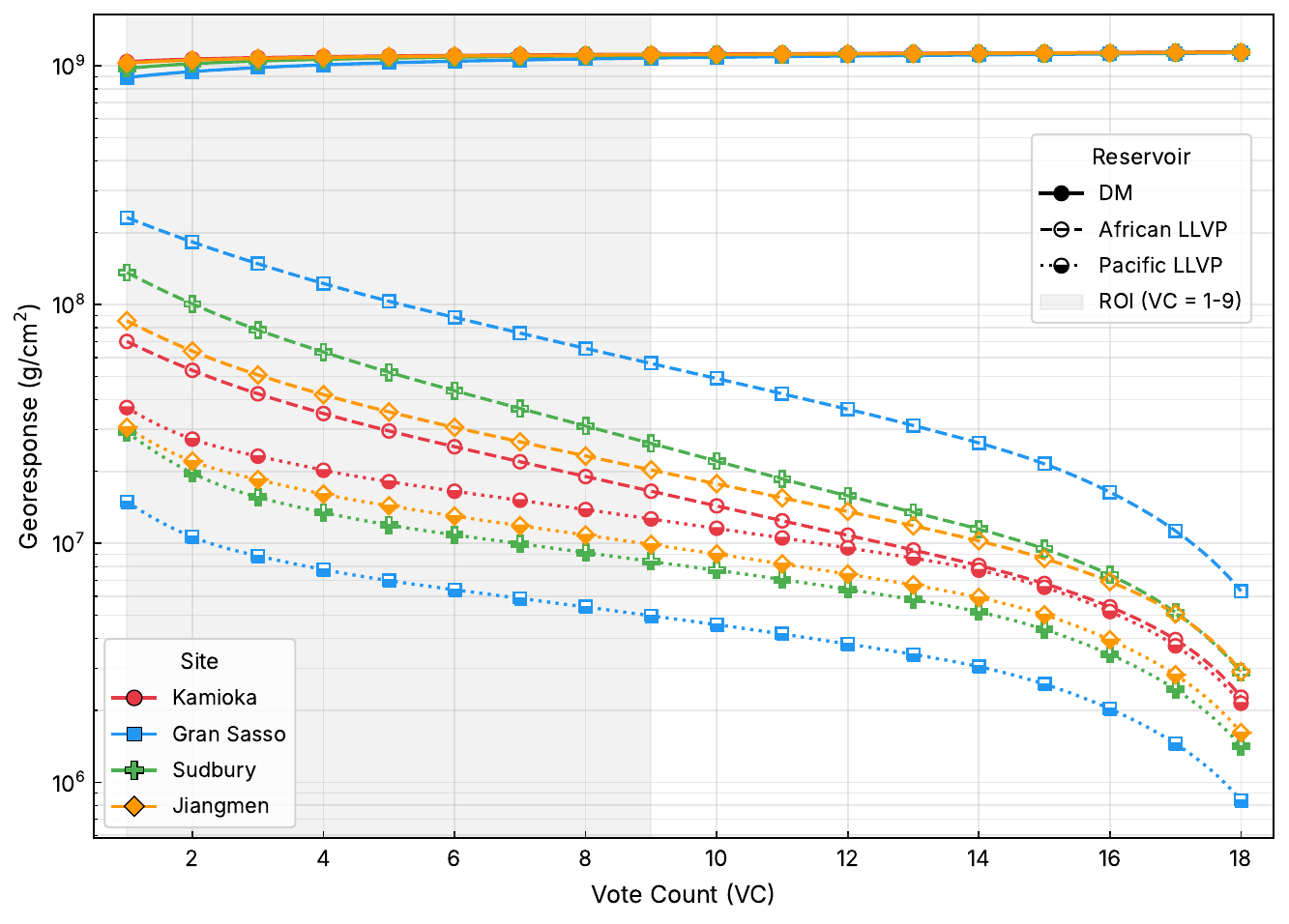}
\caption{Geo-response (g/cm$^{2}$) of the depleted mantle (DM) and Pacific and African LLVPs for our reference LLVP model (720~km and 1600~km heights, respectively) at each of the studied sites as a function of VC. The shaded band marks the VC region of interest for this analysis.}
\label{fig:geo-response_sites}
\end{figure}

The geo-response $G_{i,k}\text{(VC)}$ of the depleted mantle (DM) and Pacific and African LLVPs at each of the studied sites is shown in Figure~\ref{fig:geo-response_sites} as a function of VC for our reference LLVP model. The geo-response of the DM rises with VC and flattens toward high VC. Kamioka, Jiangmen and Sudbury sit within a few percent of each other, while Gran Sasso is lowest, especially at low VC. The geo-response of the African LLVP falls steeply with VC and Gran Sasso is most sensitive to it by a wide margin. For the Pacific LLVP, the ordering flips, as Kamioka is the most sensitive site and Gran Sasso the least. All four curves fall toward zero at high VC.

The results of the constrained and unconstrained fits for our reference LLVP model are discussed in Appendix~\ref{app:ref_llvp_model_fit}.

\section{Slow- and Fast-Velocity Anomalies (S and F)\label{sec:sf_anomalies}}

Similar to the LLVP models described in Section~\ref{sec:llvp_models}, the slow- and fast-velocity-anomaly families of models place anomalous bodies in the mantle using tomographic vote maps. Vote maps are downloaded from SubMachine~\cite{SubMachine} and \cite{Shephard_2017} at depths between 100--2800~km in slices of 100~km. A cell with $\text{VC} \geq c$ is one where at least $c$ of the 18 models agree on the anomaly. We scan $c \in \{ 1,\,3,\,5,\, \dots \, 17\}$.

To construct the slow- or fast-velocity anomalous bodies, the vote maps are node-registered on a 0.5$^{\circ}$ grid. The duplicate 360$^{\circ}$ longitude column is dropped. Each node represents a cell of $\pm0.25^{\circ}$, with half cells at the poles. 
A slice at depth $d$ represents the shell from $d - 50$~km to $d+50$~km. The 2800~km slice extends to the CMB. The shallowest shell starts at 50~km, which matches the top of the mantle (39--65~km) in the geo-response tables of our LLVP model. Densities of each shell are taken from PREM~\cite{dziew81} at the shell center. The geo-response for a volume element at distance $L$ from a detector at the surface is then calculated as in Equation~\ref{eqn:geo-response}.  

Cells within 400~km of a detector (or $3\times$ depth) and shallower than 1000~km are sub-sampled $5 \times 5$ laterally and 10$\times$ vertically for that detector. All other cells are treated as points.

The part of a shell above the lithosphere-asthenosphere boundary is considered lithosphere, already counted in $S_\text{LS}$. The boundary depth is taken from the nearest LITHO1.0 node (LID bottom)~\cite{Pasyanos2014LITHO10}. The mask can be configured on/off, and matters only between 100--200~km, where cratonic keels lie. 

African and Pacific domains are distinguished using latitude and longitude cuts. A cell is considered African if it is closer to (10$^{\circ}$S, 20$^{\circ}$E) than to (10$^{\circ}$S, 170$^{\circ}$W), otherwise Pacific. This rule reproduces the LLVP masses and geo-responses of the LLVP models to within 1--2\% for $\text{VC} \geq 3$.

We model two classes of \textbf{S}-family anomalies:
\begin{itemize}
    \item \textbf{S-a}: CMB-attached bodies. These are essentially LLVPs rebuilt from the low-velocity vote maps. Slow-velocity cells with $\text{VC} \geq c$ from the CMB up to independent heights $H_\text{Afr}$, $H_\text{Pac} \in \{ 100,\,400,\, \dots 2500\}$~km in 300~km steps, plus 2840~km (the full mantle column), with the lithosphere mask on. We scan over heights, $c$, OCR fraction, and $k$, while fitting BR, for a total of 388,800 configurations. 
    \item \textbf{S-b}: Global depth windows. All low-velocity cells with $\text{VC} \geq c$ between a top slice (100--2500~km) and a bottom slice (100--2800 km), globally, assuming a single composition: $\text{OCR} = \{ 0,\,0.23,\,0.28,\,0.33 \} + f*\text{BCC}$, with $f$ left as a fit parameter. There is no African/Pacific split and no other LLVP. Lithosphere mask can be on or off. A total of 84,240 configurations are tested. 
\end{itemize}

For \textbf{F}-family anomalies, all fast-velocity cells with $\text{VC} \geq c$ between a top slice (100--2500~km) and a bottom slice (100--2800~km), globally, are assigned a single composition: $\text{OCR} = \{ 0,\,0.07,\,0.15\} + f*\text{BCC}$. Lithosphere mask can be on or off, so that cratonic keels can be excluded. We model two classes:
\begin{itemize}
    \item \textbf{F-a}: fast-velocity anomalies residing in a DM-only mantle, with $f$ left as a free parameter.
    \item \textbf{F-b}: fast-velocity anomalies in addition to our reference LLVP model ($\text{VC}=5$, $\text{OCR}=0.28$, $k=1/10$) with $f$ and BR left as fit parameters. 
\end{itemize}
A total of 63,180 configurations are tested for each class. 

\section{Subducted Slabs (B)\label{sec:subducted_slabs}}

Subducted slabs are sections of Earth's oceanic (or sometimes continental) lithosphere that have bent and sunk downward into the mantle at convergent plate boundaries. We use age and location data from the 100 slabs cataloged in the Atlas of the Underground~\cite{vanderMeer2018} to define anomalous bodies within the mantle. Each slab is assumed to be 100~km thick and 1000~km along strike at 3.3~g/cm$^3$, with a length of 5 cm/yr times the age span of the slab. The mass of each slab is spread in a cylinder of radius 300, 500 or 800~km around the midpoint given in~\cite{vanderMeer2018}, from top to base depth. We calculate a total subducted slab mass of $9.3 \times 10^{22}$~kg (2.3\% of the mantle mass). The slab geo-response is largest at Gran Sasso ($1.9 \times 10^8$~g/cm$^2$), from the Calabria and Northern Apennine slabs, and smallest at Jiangmen ($3.1 \times 10^7$~g/cm$^2$), which is the opposite of the required pattern.

For slab composition, we scan over $\text{OCR} \in \{ 0.05,\,0.07,\,0.10\} + \text{GLOSS sediment} \in \{ 0,\,0.5,\,1 \}\% + f*\text{BCC}$, with $f$ left as a fit parameter. BCC is applied either to the 22 collision slabs or to all slabs. We consider the surrounding mantle to be DM-only or DM + the reference LLVPs. A total of 972 configurations per background are tested. 

\section{Lithosphere Components (L)\label{sec:lithosphere_components}}

In the \textbf{L}-family of models, we do not define anomalous bodies in the mantle, but instead multiply each published lithosphere component listed in Table~\ref{tab:ls_components} by a factor of 0--3 at all four sites at once. We use the reference LLVP model for this study, with $\text{VC}=1$--18 and BR fitted, giving a total of 17,496 configurations. 

\begin{table}
\caption{Lithosphere components used in the \textbf{L} family including the continental lithospheric mantle (CLM). Far-field crust and CLM values at Kamioka are from~\cite{huang_etal_2013}; local crust values are derived here (see main text). Values at Gran Sasso are from~\cite{agostini_etal_2020}, with local crust values from~\cite{Coltorti_2011} and far-field crust from~\cite{huang_etal_2013}. Values at Sudbury and Jiangmen are from~\cite{strati_etal_2017} and \cite{Gao2020JULOC}, respectively. All values are given in NIU.} 
\setlength{\tabcolsep}{3pt}
\begin{tabular}{l c c c c}
\hline\noalign{\smallskip}
Component & Kamioka & Gran Sasso & Sudbury & Jiangmen \\
\hline\noalign{\smallskip}
Upper crust (local) & 10.0 & 5.84 & 12.9 & 24.5 \\
Middle crust (local) & 2.4 & 0 & 1.2 & 2.1 \\
Lower crust (local) & 0.27 & 0.57 & 0.6 & 0.15 \\
Sediments (local) & 0.21 & 2.79 & 0 & 0 \\
Far-field crust & 7.31 & 14.0 & 15.2 & 9.8 \\ 
CLM & 1.6 & 2.3 & 2.1 & 2.1 \\
\hline\noalign{\smallskip}
\end{tabular}
\label{tab:ls_components}
\end{table}

At Kamioka we use the published value of the far-field crust from~\cite{huang_etal_2013}, $7.31^{+1.5}_{-1.2}$~NIU. No local crust model is consistent with the Huang et al.\ far-field value. The regional model in~\cite{ENOMOTO2007147} uses different crustal abundances. The local layers at Kamioka are therefore derived here. We use whole-crust layers (20.2~NIU) from \cite{huang_etal_2013} and subtract the far-field crust value (7.31~NIU), with the remaining 12.9~NIU shared among the layers in proportion to their CRUST1.0 signal within 500~km of the detector. 

The upper crust is the only strong lever since Jiangmen's local upper crust is 2.5$\times$ that of Kamioka. As the upper crust is scaled up, the best fits in this family compensate by lowering the mantle contribution: the preferred DM abundance falls to $a_\text{DM}\text{(U)} = 8$, 6.5 and 4.7~ppb (Th/U = 2.5) and the best-fit BR drops to 12.6, 9.3 and 4.7\% for 1.25$\times$, 1.5$\times$ and 2$\times$, respectively.

\section{Local Site-Specific Models\label{sec:local_models}}

We run three classes of local site-specific models by applying rules that use lateral distance cuts or act on individual detectors:
\begin{itemize}
    \item \textbf{C-S}: slow-velocity anomalies within $R \in \{ 300,\,600,\,1000,\,2000\}$~km of any detector and outside the reference LLVPs are added to the reference LLVPs with $\text{OCR} = \{ 0,\,0.28 \} + f*\text{BCC}$.
    \item \textbf{C-F}: the same for fast-velocity anomalies, with $\text{OCR} \in \{ 0,\,0.07 \}$.
    \item \textbf{C-J}: a site-specific offset at Jiangmen. A free additional mantle signal (0--40~NIU) is applied at Jiangmen only, on top of the reference LLVPs. This is similar to the leave-one-out test in Section~\ref{sec:homogeneous_fit_withpulls}, but on an LLVP background instead of a uniform one.
\end{itemize}
A total of 136,512 (\textbf{C-S}), 166,248 (\textbf{C-F}), and 18 (\textbf{C-J}) configurations are tested.

\section{Likelihood Framework\label{sec:likelihood_framework}}

Every fit minimizes the same total $\chi^{2}$ using the same four measured rates and published lithosphere predictions from Table~\ref{tab:geonu_rates_sites}:

\begin{widetext}
\begin{align}
    \chi^2_\text{total} = & \sum_{i} \chi^2_{i}(S_\text{pred}) + \sum_{i} \delta^2_{i} + \chi^2_{\mathrm{geo}} +\chi^2_{\mathrm{prior}} \\
    = & \sum_{i} \left[ \left(\frac{S_{\mathrm{mantle},i} + 
    S_{\mathrm{LS},i}(\delta_i) - S_{\mathrm{obs},i}}{\sigma_i^{\pm}}\right)^2 + \delta_i^2 \right] + w \cdot \left( \frac{\text{BR} - \text{BR}^\text{nom}_\text{pred}(\text{VC})}{\sigma_\text{band}(\text{VC})} \right) ^2 + w_{\text{VC}} \cdot \left( \frac{\text{VC} - \text{VC}^\text{prior}}{\sigma_{\text{VC},\,\text{prior}}} \right) ^2,
    \label{eqn:total_chi2}
\end{align}
\end{widetext}
where $\delta_i$ are the lithosphere nuisance parameters at each site and $\sigma_i^{\pm}$ is the upper or lower experimental error, depending on the sign of the residual (split Gaussian). The geochemical constraint and tomographic prior can be turned on ($w,\,w_\text{VC}=1$) or off ($w,\,w_\text{VC}=0$) independently. Note that the tomographic prior only applies to families of models that contain LLVPs.

As before, each $\delta_i$ is a dimensionless pull on that site's lithosphere prediction: $S_\text{LS,eff}(i) = S_\text{LS,nom}(i)(1+u_i \delta_i)$, where $u_i$ is the site's fractional lithosphere-model uncertainty, switching smoothly between the downward and upward fractional lithosphere uncertainty. The $\delta^2_i$ term is a standard Gaussian penalty: each nuisance costs exactly 1 unit of $\chi^2$ per sigma it is pulled away from its nominal (zero) value. This is the usual ``pull term" construction for a systematic uncertainty with a well-defined prior width. All four $\delta_i$ are genuine free parameters of the same Minuit object as the signal parameters. 

Since each $\delta_i$ couples to one site only, the minimization over $\delta_i$ is exact and done once: $\chi^{2}_i(S)$ is tabulated at each site for $S = -40$ to 160~NIU in steps of 0.01~NIU, with $\delta_i$ from -5 to 5 in steps of 0.005. The fit then minimizes over the composition parameters on a grid. BR uses a 0.1\% step, while $f$ uses a 0.0025 step when fitted alone. When BR and $f$ are fitted together, a $0.5\% \times 0.01$ grid is refined locally on a $21 \times 21$ grid.

We count every term in Equation~\ref{eqn:total_chi2} as a measurement and every fitted quantity as a parameter. Each nuisance parameter therefore cancels its penalty term, and the geochemical constraint and VC prior each add one degree of freedom.

\section{Common scan ranges and plausibility\label{sec:scan_ranges}} 

For all families, DM abundances are scanned over $a_\text{DM}(\text{U}) \in \{ 3.2,\, 4.7,\, 6.5,\, 8,\, 10,\, 12 \}$~ppb, each with $\text{Th/U} \in \{ 2.5,\,2.7,\,2.9 \}$ for a total of 18 combinations. This range spans the depleted-mantle estimates of \cite{workman_hart_2005} (3.2 ppb U, 7.9 ppb Th), \cite{salters_stracke_2004} (4.7 ppb U, 13.7 ppb Th), \cite{Arevalo2010Chemical,huang_etal_2013} (8 ppb U, 22 ppb Th), including uncertainties. Each configuration is fitted with $w \in \{ 0,\,1 \}$. Configurations with a free $f$ are also fitted both with $f$ unrestricted (0--1) and with $f \leq 0.15$.

A fit is considered ``plausible" when all of the following hold:
\begin{itemize}
    \item The geochemical constraint is on ($w = 1$), and the tomographic prior is included ($w_\text{VC} = 1$), where defined.
    \item $f \leq 0.15$. This bounds the BCC fraction of additional bodies and is tied to the inventory of recycled continental crust constrained by literature.
    \item MORB-consistent: no BCC ($f > 0.01$) in slow-velocity bodies that extend above 350~km depth (top slice at 300 km or shallower). Global mean MORB has $\text{U} = 0.119 \pm 0.013$~ppm~\cite{Gale2013}, implying a source of roughly 3--12~ppb U. This rule is applied to the global slow-velocity (\textbf{S}) and the local slow-velocity families (\textbf{C-S}).
    \item Lithosphere components scaled by 0.75--1.25, roughly the published upper-crust uncertainty.
\end{itemize}

\section{Results\label{sec:results}}

We study the figure of merit, defined in Equation~\ref{eqn:FOM} as $S_\text{mantle}\text{(Jiangmen)} - S_\text{mantle}\text{(Kamioka)}$, for all families, model configurations, and fits totaling $\sim$3.0 million. For the lithosphere family (\textbf{L}), the lithosphere shift is added: $\Delta S_\text{eff} = \Delta S + (\text{shift(Jiangmen)} - \text{shift(Kamioka)})$. 

The data ``best estimate" value comes from the same profiled site likelihoods used in the fits (with split-Gaussian measurement errors and profiled lithosphere nuisance parameters), minimized along $\Delta S$, with $1\sigma$ intervals at $\Delta \chi^2 = 1$:
\begin{itemize}
    \item $S_\text{mantle}\text{(Kamioka)}$: 8.0~NIU (1$\sigma$: 1.3--14.1).
    \item $S_\text{mantle}\text{(Jiangmen)}$: 32.6~NIU (1$\sigma$: 20.3--44.7).
    \item $\Delta S_\text{data} = 24.6 \pm 13.8$~NIU (1$\sigma$: 10.9--38.4), with $\Delta \chi^2 = 3.2$ at $\Delta S = 0$ (1.8$\sigma$) and a one-sided $P(\Delta S_\text{data} \leq 5\text{ NIU}) = 0.077$.
\end{itemize}
A Monte Carlo of the observed rates minus the lithosphere predictions, with split-Gaussian errors, gives 24.7~NIU, in agreement. This value also agrees with that calculated from the naive mantle model (Section~\ref{sec:naive_mantle_model}). In the $S_\text{mantle}\text{(Kamioka)}$-$S_\text{mantle}\text{(Jiangmen)}$ plane, the data region is drawn from the profiled site likelihoods at $\Delta \chi^2 =2.30$ and 6.18. The best estimate is (8.0, 32.6)~NIU. 

The reference used here is the uniform mantle fit from Section~\ref{sec:homogeneous_fit_withpulls}: one free mantle signal common to all four sites, plus the four lithospheric nuisance parameters. Its best fit is $S_\text{mantle} = 14.8$~NIU at every site, with $\chi^2 = 3.78$ and $\Delta S = 0$ by construction. The best-fit improvement for each family is quoted as $\Delta \chi^2 = \chi^2_\text{hom} - \chi^2_\text{best}$. Positive values are better fits than the uniform mantle. No family is measured against the reference LLVP fit. 

A summary of the results is presented in Table~\ref{tab:results_summary}. LLVP models (\textbf{H}) result in $\Delta S = -0.5$ to 1.2~NIU for plausible fits. Slow-velocity-anomaly models produce $\Delta S = -2.7$ to 3.3~NIU (\textbf{S-a}) or $\Delta S = -1.1$ to 1.8~NIU (\textbf{S-b}) for plausible fits. Unconstrained fits in the \textbf{S} family reach up to 21~NIU, but only with $f=0.98$ (nearly pure BCC) at a depth of 200--300~km, which is ruled out by MORB studies. Fast-velocity-anomaly models produce $\Delta S = -4.1$ to 2.1~NIU (\textbf{F-a}) or $\Delta S = -3.2$ to 1.9~NIU (\textbf{f-b}) for plausible fits. Unconstrained fits in this family reach up to 6.2~NIU with pure BCC at a depth of 1000--1500~km. Plausible fits employing subducted slabs (\textbf{B}) result in $\Delta S = -2.7$ to 0.9~NIU. Scaling individual lithosphere components, particularly the upper crust, by 1.25$\times$, 1.5$\times$, and 2$\times$ results in $\Delta S$ as high as 4.3, 7.7, and 14.7~NIU, respectively. The latter two results are outside the defined plausibility range. Local models in the \textbf{C-S} family can produce $\Delta S = 24$~NIU, but only with $f \approx 1$ in the shallow mantle near Jiangmen. Local fits that are MORB-consistent give $\Delta S \leq 4.0$~NIU.

\begin{table*}[t]
\caption{Summary of results sorted by family. All $\Delta S$ values are given in NIU. The column ``Max $\Delta S$ near best" is the largest $\Delta S$ among plausible fits within $\Delta \chi^2 = 1$ of the family's best plausible fit.} 
\setlength{\tabcolsep}{3pt}
\begin{tabular}{l c c c c c c c c}
\hline\noalign{\smallskip}
Family & Config. & Best plaus. & $\Delta \chi^2$ vs.\ & $\Delta S$ at best& Plaus. $\Delta S$ & Max $\Delta S$ & Best un- & Unconstr. \\
& & $\chi^2$ & homogen. & plaus. & range & near best & constr. $\chi^2$ & $\Delta S$ range \\
\hline\noalign{\smallskip}
Uniform mantle (ref.) & 1 & 3.78 & -- & 0 & 0 & 0 & 3.78 & 0 \\
\hline\noalign{\smallskip}
\textbf{H}: LLVP Models & 46,656 & 3.05 & +0.74 & 0.7 & -0.5--1.2 & 0.9 & 2.95 & -0.7--1.5 \\
\textbf{S-a}: slow, CMB-attached bodies & 388,800 & 2.86 & +0.92 & 1.5 & -2.7--3.3 &  1.9 & 2.60 & -3.3--3.2 \\
\textbf{S-b}: slow, global depth windows & 84,240 & 3.36 & +0.42 & 1.7 & -1.1--1.8 &  1.8 & 1.70 & -2.1--21.2 \\
\textbf{F-a}: fast, DM only & 63,180 & 3.33 & +0.45 & 1.5 & -4.1--2.1 & 2.0 & 2.93 & -6.3--6.2 \\
\textbf{F-b}: fast + reference LLVPs & 63,180 & 2.85 & +0.93 & 1.8 & -3.2--1.9 & 1.9 & 2.26 & 0.1--4.9 \\
\textbf{B}: slabs, DM only & 972 & 4.02 & -0.23 & -2.2 & -2.7--0.6 & 0.4 & 3.37 & -2.9--1.2 \\
\textbf{B+}: slabs + reference LLVPs & 972 & 3.07 & +0.71 & 0.7 & -2.2--0.9 & 0.9 & 3.06 & 0.1--1.1 \\
\textbf{L}: lithosphere components ($\Delta S_\text{eff}$) & 17,496 & 2.07 & +1.71 &  4.3 & -3.6--4.6 & 4.4 & 1.26 & -14.3--29.1 \\
\textbf{C-S}: local slow anomalies & 136,512 & 2.48 & +1.30 & 4.0 & -0.6--4.0 & 4.0 &  0.03 & 0.2--23.7 \\
\textbf{C-F}: local fast anomalies & 166,248 & 2.99 & +0.80 & 1.0 & -5.8--1.0 & 1.0 &  2.81 & 0.1--2.5 \\
\textbf{C-J}: site-specific offset at Jiangmen & 18 & 0.17 & +3.61 & 22.7 & 18.9--29.7 &  25.0 & 0.02 & 20.7--24.4 \\
\hline\noalign{\smallskip}
\textbf{Total} (scanned configurations) & \textbf{968,274} & & & & & & \\
\hline\noalign{\smallskip}
\end{tabular}
\label{tab:results_summary}
\end{table*}

\begin{figure*}
    \centering
    \includegraphics[trim = 0mm 0mm 0mm 0mm, width=0.95\linewidth]{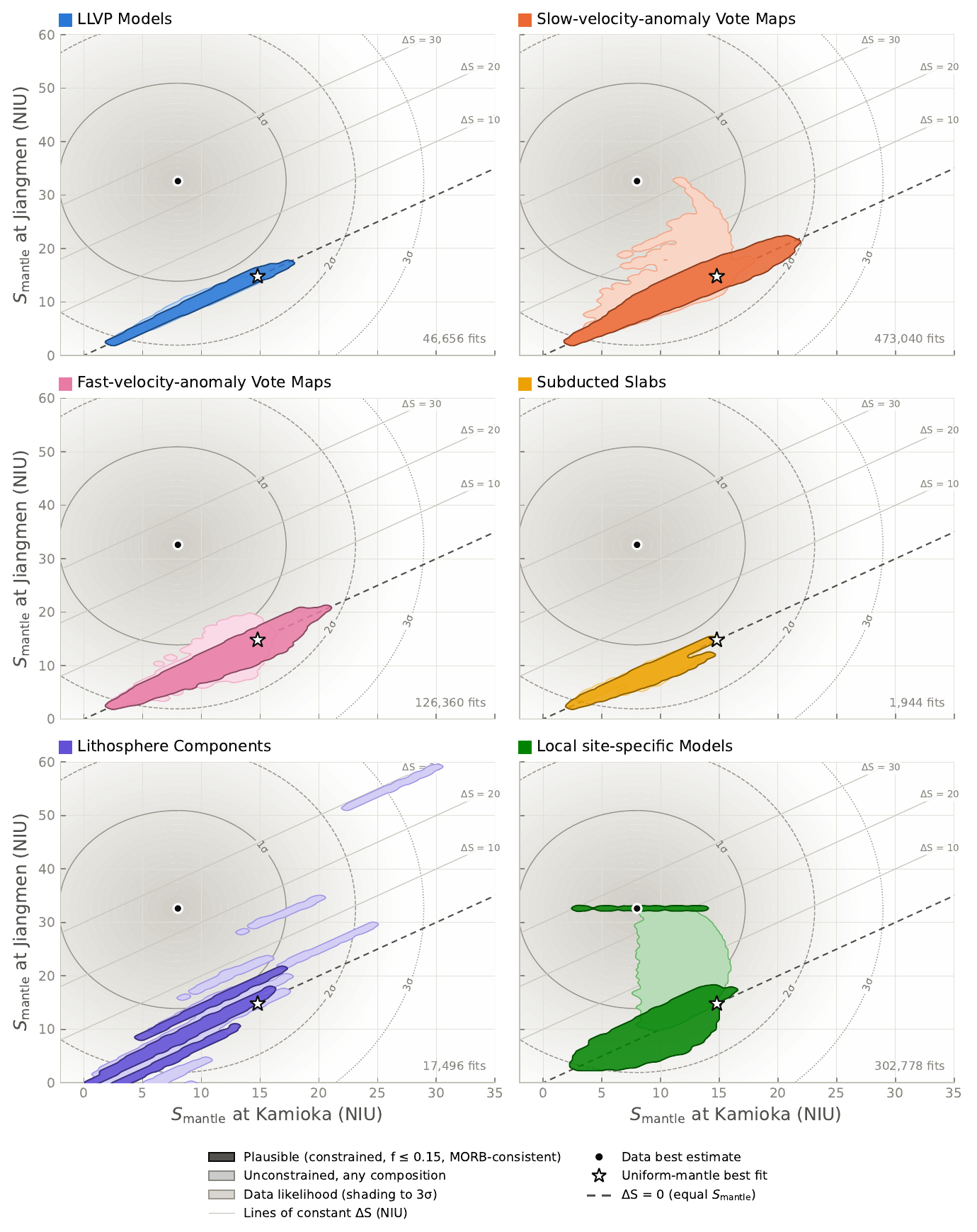} 
    \caption{$S_\text{mantle}\text{(Jiangmen)}$ (NIU) vs.\ $S_\text{mantle}\text{(Kamioka)}$ (NIU) for all model configurations and fits ($\sim$3.0 million) studied here, disaggregated into one panel per family. Solid regions with dark outlines are plausible fits (constrained, $f \leq 0.15$, MORB-consistent). Pale regions are unconstrained fits of any composition. Regions are drawn from the fitted points, widened by about 0.5~NIU so that thin features stay visible. Grey shading is the data likelihood out to 3$\sigma$, with labeled 1$\sigma$, 2$\sigma$, and 3$\sigma$ contours around the data best estimate (black dot at 8.0, 32.6~NIU). The dashed line is $\Delta S = 0$, and thin diagonal lines mark $\Delta S$ = 10, 20, and 30~NIU. The star represents the uniform mantle best fit at (14.8, 14.8)~NIU.}
    \label{fig:results_separate}
\end{figure*}

\begin{figure*}
    \centering
    \includegraphics[trim = 0mm 0mm 0mm 0mm, width=0.75\linewidth]{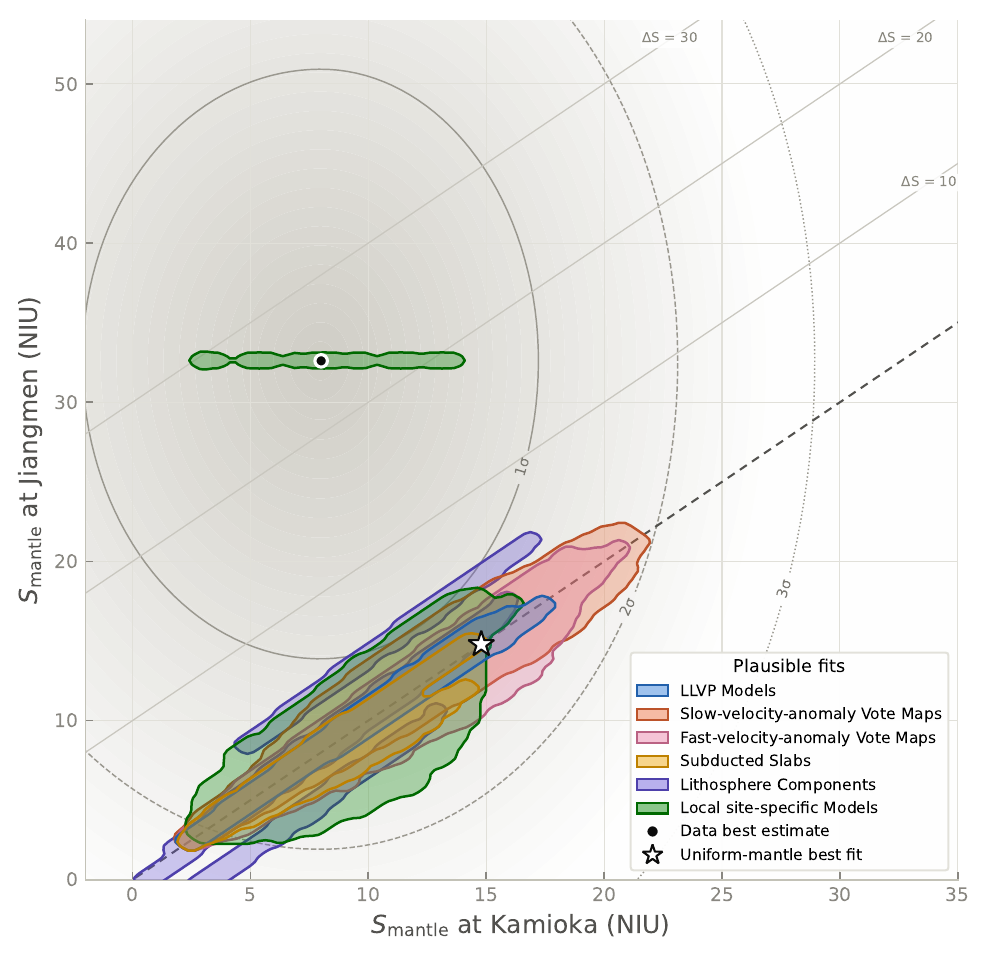} 
    \caption{Summary of Figure~\ref{fig:results_separate}: $S_\text{mantle}\text{(Jiangmen)}$~(NIU) vs.\ $S_\text{mantle}\text{(Kamioka)}$~(NIU), with the plausible regions of all six families overlaid on one panel (constrained, $f \leq 0.15$, MORB-consistent). Colors are as in Figures~\ref{fig:results_separate} and \ref{fig:results_bar}. Regions are drawn from a total of $\sim$3.0 million points, widened by 0.5~NIU so that thin features stay visible. Grey shading is the data likelihood out to 3$\sigma$, with labeled 1$\sigma$, 2$\sigma$, and 3$\sigma$ contours around the data best estimate (black dot at 8.0, 32.6~NIU). The dashed line is $\Delta S = 0$, and thin diagonal lines mark $\Delta S = 10$, 20, and 30~NIU. The star represents the uniform mantle best fit (14.8, 14.8)~NIU. All universal families cluster along $\Delta S = 0$. Only the site-specific offset at Jiangmen reaches the data region.} 
    \label{fig:results_summary}
\end{figure*}

\begin{figure*}
    \centering
    \includegraphics[trim = 0mm 0mm 0mm 0mm, width=0.99\linewidth]{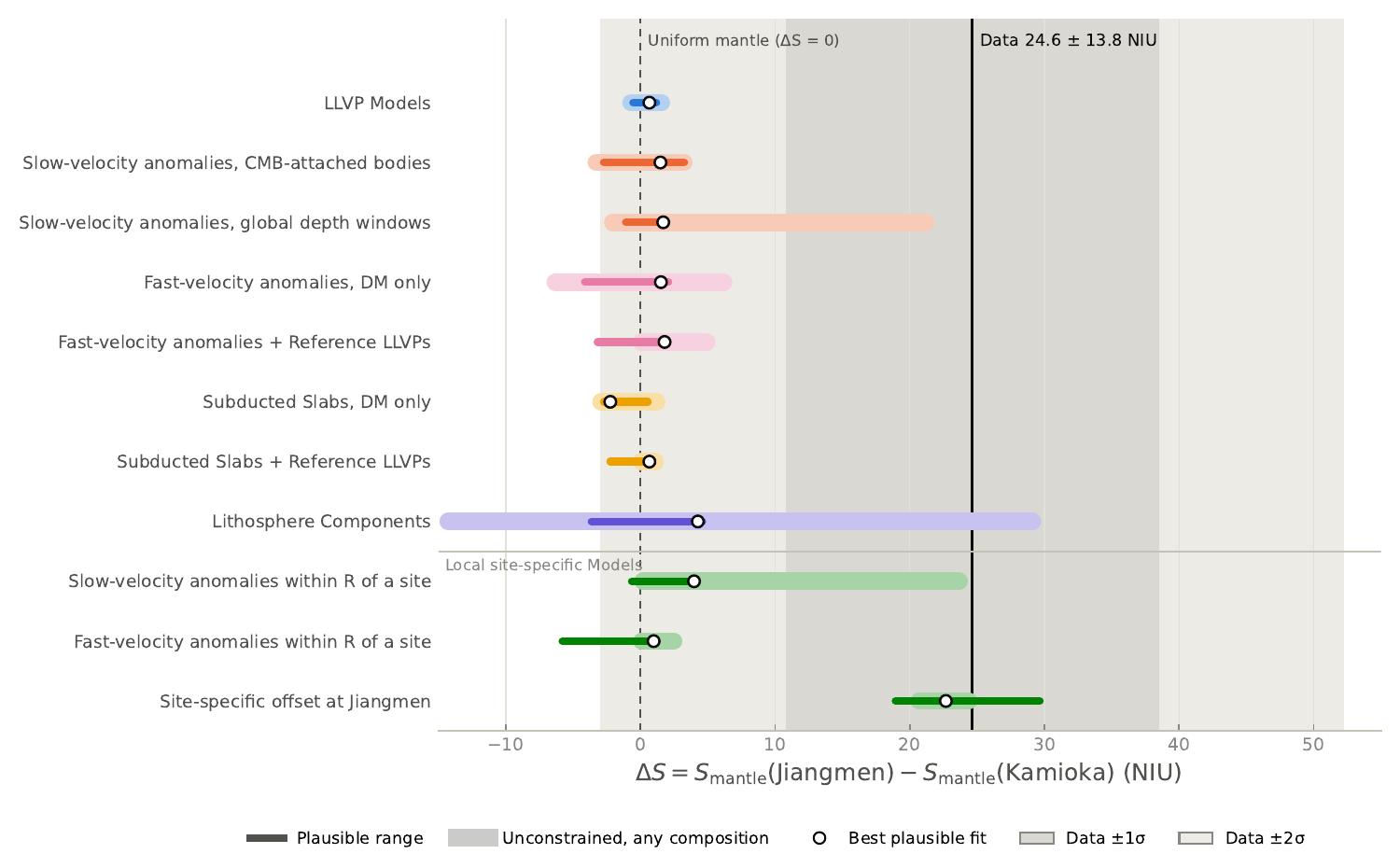} 
    \caption{$\Delta S = S_\text{mantle}\text{(Jiangmen)} - S_\text{mantle}\text{(Kamioka)}$~(NIU) for all model configurations and fits ($\sim$3.0 million) studied here, grouped by family. Dark bars show the plausible range for each family, light bars show the unconstrained range, and circles the best plausible fit. The dashed line is the uniform mantle best fit ($\Delta S = 0$). Shaded bands show the data best estimate $\Delta S_\text{data} = 24.6 \pm 13.8$~NIU at 1$\sigma$ and 2$\sigma$. Among plausible fits, only the site-specific offset at Jiangmen reaches the 1$\sigma$ band.}
    \label{fig:results_bar}
  \end{figure*}

In summary, the non-local families (\textbf{H}, \textbf{S}, \textbf{F}, \textbf{B}, \textbf{L}) give $\Delta S \leq 4.0$~NIU, against $24.6 \pm 13.8$~NIU required by the data, with plausible compositions and scaling. The best fits in these families improve on the uniform mantle by at most $\Delta \chi^2 = 1.7$. Only local, site-specific rules or a free offset at Jiangmen reach the data. Figures~\ref{fig:results_separate} and \ref{fig:results_summary} summarize the results of all $\sim$3.0 million fits in the Kamioka--Jiangmen mantle-signal plane, per family. Figure~\ref{fig:results_bar} summarizes the same set of results per family, but looking at $\Delta S$ ranges instead.

\section{Statistical context and limitations\label{sec:limitations}}


The pull pattern (Kamioka low, Jiangmen high) is consistent with statistical scatter. At the unconstrained reference LLVP fit ($\chi2 = 3.05$, Appendix~\ref{app:ref_llvp_model_fit}), the residuals, including the lithosphere uncertainty, are $-0.66\sigma$, $-0.11\sigma$, $-0.08\sigma$, and $+1.61\sigma$ at Kamioka, Gran Sasso, Sudbury and Jiangmen, respectively. In 1000 pseudo-experiments generated from this fit and refitted, $p(\chi^2 \geq 3.05) = 0.37$, and at least one site lands as far out as Jiangmen 39\% of the time. The specific ``Kamioka low and Jiangmen high” combination has $p = 0.016$, but it was chosen after seeing the data. The pseudo-experiments assume the model is correct and keep the VC prior fixed.

The Th/U mass balance in the reference LLVP fit requires approximately 1.9 continental-crust masses ($M_\text{CC} = 2.2 \times 10^{22}$~kg) of BCC in the LLVPs, assuming BCC abundances from Table~\ref{tab:abundances}. This is at or above the published upper bound for recycled continental crust. About 1.1 $M_\text{CC}$ has returned to the mantle since 3~Ga at about 3~km$^3$/yr~\cite{Scholl2009} or up to about 1.9 $M_\text{CC}$ at present-day loss rates~\cite{Stern2011SubductionER}, assuming no relamination.

The following limitations of this analysis are important to mention:
\begin{itemize}
    \item Vote-map bodies use PREM density for every composition.
    \item The African/Pacific split is a geometric nearest-center rule. 
    \item Slab geometry is schematic: fixed thickness, convergence rate, and along-strike length.
    \item The lithosphere components come from different site-specific models, and the local split at Kamioka is computed here.
    \item The mantle tables start at a fixed depth of about 40--65 km, not at the lithosphere base. At typical DM abundances, they double-count about 0.3, 0.25, 0.6 and 0.2~NIU of lithospheric mantle at Kamioka, Gran Sasso, Sudbury and Jiangmen, respectively.
\end{itemize}

\section{Summary and Conclusions\label{sec:conclusions}}

Our global fit to geo-neutrino measurements at four underground sites (Kamioka, Gran Sasso, Sudbury, Jiangmen) provides a constraint on the quantity and distribution of radioactivity in the Earth's mantle. By combining observed signal rates with site-specific predictions of the lithospheric contribution and assuming a homogeneous distribution of radioactivity in the mantle, we find a geo-neutrino rate of $S_\text{mantle} = 14.8 \pm 4.5$~NIU, when allowing the lithosphere predictions to vary within their uncertainties. This uniform mantle signal corresponds to radiogenic heating of $18.5 \pm 5.6$~TW, assuming Th/U$ = 3.77$ and K/U$ = 13,000$ as the ratios of the mass-fraction abundances of the main heat-producing elements. 


The data provide modest but specific motivation for considering a laterally heterogeneous mantle. Subtracting the predicted lithosphere signal from each measured rate gives naive mantle signals that differ markedly between sites, from 8.0~NIU at Kamioka to 32.6~NIU at Jiangmen. A uniform mantle fits the data acceptably overall ($\chisq/\ndof = 3.78/3$), but there is tension between the inferred mantle signal difference at Jiangmen and Kamioka ($\Delta S = 24.6 \pm 13.8$~NIU, non-zero at $\Delta \chi^2 = 3.2$, or 1.8$\sigma$). 

The two sites are only $\sim$2,900 km apart (comparable to the depth of the CMB) and see largely overlapping mantle volumes, so a uniform mantle, or any large-scale structure, should give them nearly the same signal. A real difference of this size would therefore imply heat-producing material concentrated much closer to one detector than the other. The pattern was identified after seeing the data and has modest significance, so it does not by itself establish heterogeneity. It does define a clear, testable target: whether any geochemically and geophysically plausible distribution of U and Th can produce a Jiangmen–Kamioka contrast of this size. We therefore use $\Delta S$ as the figure of merit for the heterogeneous models examined here.

We explore six families of models to describe the heterogeneous spatial distribution of radiogenic elements in the Earth’s mantle. Each model contains anomalous bodies or reservoirs carrying an excess of heat-producing elements over a depleted mantle (DM) background, with total inventory set by bulk-silicate-Earth mass balance and by a geochemical Th/U constraint. A total of 968,274 individual model configurations are defined by varying the parameter space, and then tested against the figure of merit $\Delta S$.

The results of $\sim$3 million distinct fits show that no universal mantle model with plausible composition produces more than $\sim$3--4~NIU of Jiangmen$-$Kamioka mantle-signal difference, while the required difference from data is $\sim$25~NIU. The upper crust is the most effective lever ($\Delta S_\text{eff} = 4$~NIU within plausible scaling), and in practice points to Jiangmen's local crust model. Only local or site-specific rules reach the data, while the scatter itself is statistically ordinary. This finding suggests that the tension between the rates at Kamioka and Jiangmen is due to a local effect or statistical scatter, rather than large-scale mantle structure. Revised lithosphere predictions and additional site exposures would establish whether the tension persists.

In summary, these results show that current geo-neutrino measurements begin to probe lateral heterogeneity in the mantle, and that the present tension between results at Kamioka and at Jiangmen is not explained by plausible variations of the lithosphere or by any of the large-scale mantle structures examined here, including LLVPs, seismic velocity anomalies, and subducted slabs. Because Kamioka and Jiangmen sample largely overlapping mantle volumes, any deep, laterally extended source makes similar contributions to both sites. A contrast of the observed size thus requires additional heat-producing material concentrated much closer to one of the sites.

\appendix

\section{Point Source Fits\label{app:point_source_fits}}

To further probe heterogeneity in the mantle, while remaining independent of any tomographic or geochemical assumptions, we fit an alternative, physically distinct model. Here we replace the uniform mantle with a single point source inside the Earth whose signal falls off as the inverse square of the distance to each site:
\begin{equation}
    S_{\text{mantle, }i}= \frac{L}{d_{i}^2},
\end{equation}
where $L$ is the source strength (NIU\,$\cdot$\,km$^2$) and $d_i$ is the 3D Euclidean distance from the source to site $i$, each expressed in an Earth-centered Cartesian frame built from (latitude, longitude, radius). The free parameters are: latitude, longitude, the radius from the Earth's center, and $L$. 

We perform three variants of this fit. The first two variants keep the position of the point source free, but have as many parameters as data points. Since these variants have 0 effective degrees of freedom, they are used only as plausibility/localization checks. 

\subsection{Free Point Source}

Here we keep the position of the point source free, giving four physical parameters plus four lithosphere nuisances against four data points (combined U+Th per site). This is exactly saturated (0 effective DOF). A perfect or near-perfect $\chi^2$ is therefore expected and should not be read as evidence of anything on its own. The value of this exercise is different: it shows where such a source would have to sit to explain the data, which can be compared to known LLVP locations as a plausibility check.

Because the predicted mantle rate is linear in $L$ at any fixed location, the optimal $L$ at fixed (latitude, longitude, $r$) has a closed-form solution. We first run a coarse ($10^{\circ} \times 10^{\circ} \times 50$~km) scan, then refine with a full Minuit fit (position + $L$ + all nuisances free).
  
\begin{figure} [H]
\centering
\includegraphics[trim = 0mm 0mm 0mm 0mm, clip, scale=0.3]{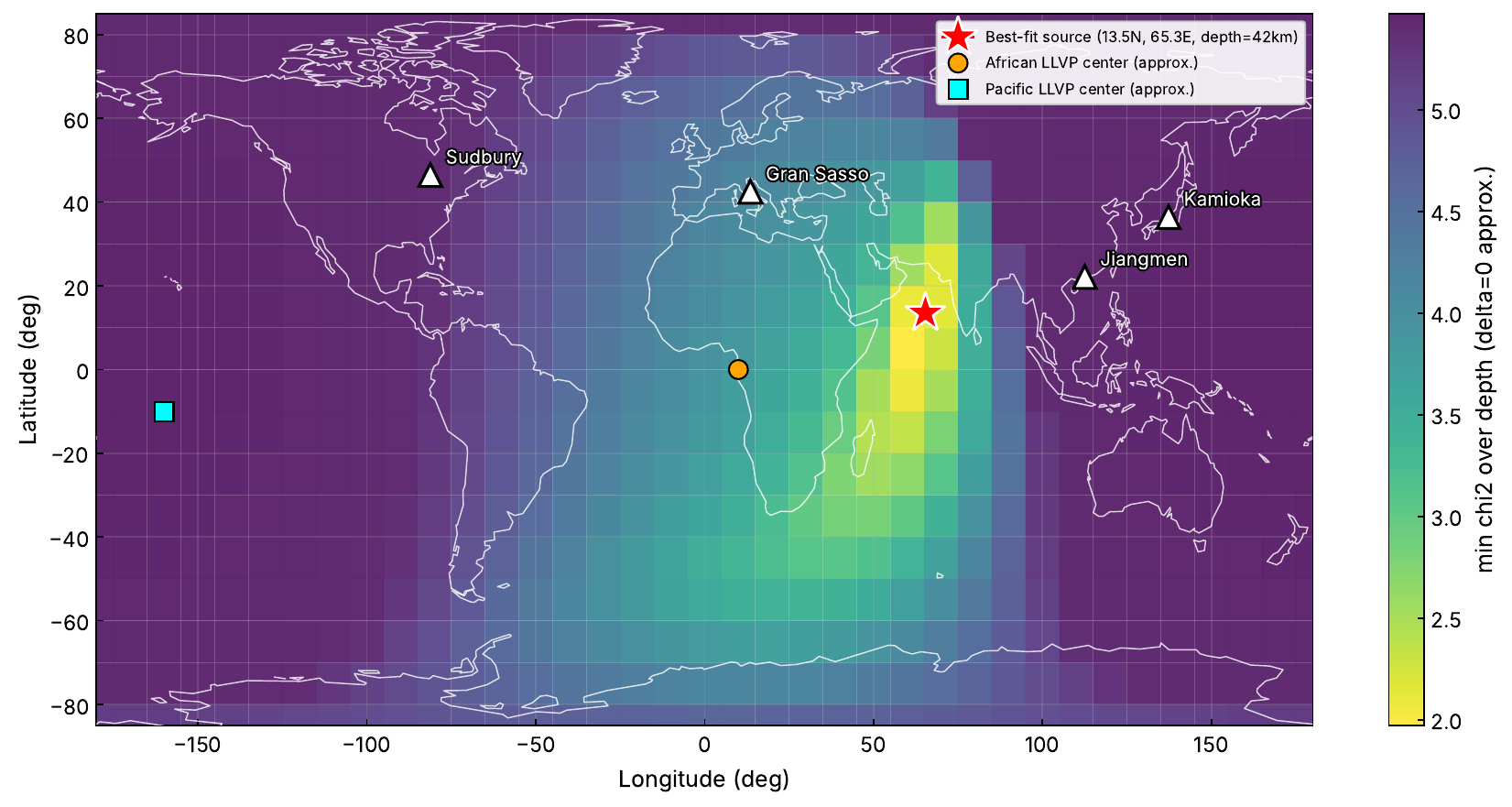}
\caption{Skymap of best fit results of point source localization with $L$ and $r$ free, marginalized over depth. The best-fit parameters are obtained at 
latitude = $13.51^{\circ} \pm  39.57^{\circ}$, longitude = $65.35^{\circ} \pm 13.65^{\circ}$, $r = (6329.0 \pm 3292.8$)~km (42.0 km depth), and $L = (6.339 \pm 6.079) \times 10^{8}$~NIU$\cdot$km$^2$, with $\chi^2 = 1.1798$.}
\label{fig:point_source_free}
\end{figure}

The best-fit parameters are obtained at latitude = $13.51^{\circ} \pm  39.57^{\circ}$, longitude = $65.35^{\circ} \pm 13.65^{\circ}$, $r = 6329.0 \pm 3292.8$~km (42.0 km depth), and $L = (6.339 \pm 6.079) \times 10^{8}$~NIU$\cdot$km$^2$, with $\chi^2 = 1.1798$ (see Figure~\ref{fig:point_source_free}). The lithosphere nuisance pulls are $-0.520 \sigma, +0.078 \sigma, +0.329 \sigma, and +0.236 \sigma$ at Kamioka, Gran Sasso, Sudbury, and Jiangmen, respectively. Note that the large parameter errors, relative to their physical range, are expected since there are 0 effective DOF. This means that the location is only weakly identified and is not a precision measurement. A robustness check for different starting parameters shows that a clear global minimum exists and is repeatably found (11/12 times), despite the large parameter errors. The degeneracy is in the size of the confidence region, and not due to the absence of a well-defined best point. 

The best-fit depth of 42.0 km is shallow (crustal) and not near the core-mantle boundary ($\sim$2891~km) where real LLVPs sit. This is an expected consequence of 0 effective DOF's. An unconstrained fit
prefers a shallow depth for the point source because it offers the largest achievable dynamic range in $1/d^2$ across the four sites for the smallest required $L$. This result should not be read as evidence for a shallow source. It reflects the flexibility of an under-constrained fit, not a physical preference.

\subsection{Mantle-Restricted Point Source}

As a complementary, more physically meaningful variant, we restrict the radius to the mantle below the crust ($>35$~km depth) and above the core-mantle boundary (2891~km depth). This still has 0 effective DOF's and the same interpretive caveats apply, but the resulting location is at least confined to where LLVPs could plausibly exist.

The best-fit parameters with this additional constraint are obtained at latitude = $13.59^{\circ} \pm  39.39^{\circ}$, longitude = $65.36^{\circ} \pm 13.60^{\circ}$, and a depth of ($55.3 \pm 2813.3$)~km, with $\chi^2 = 1.1799$, which is the same as for the unrestricted fit. 

The free-$L$, free-$r$ point-source fits in these sections have 0 effective degrees of freedom. Their low $\chi^2$ values are expected by construction and are not evidence of a real point source within the Earth. The useful result of these exercises is the best-fit location itself as a plausibility check. Here, the unconstrained best fit sits at an implausibly shallow depth far from any known LLVP. 

\subsection{Fixed-depth Point Source}

The third variant fixes the depth of the point source to 2891~km (at the CMB) and allows $L$ to float freely. The CMB is the natural place to anchor a point source serving as an LLVP proxy, since that is where real LLVPs actually sit. The free parameters are then latitude, longitude, and $L$: 3 physical parameters against 4 data points, giving $\ndof = 1$ and a genuine $p$-value, but now anchored to a physically motivated assumption on location.  

\begin{figure} [H]
\centering
\includegraphics[trim = 0mm 0mm 0mm 0mm, clip, scale=0.3]{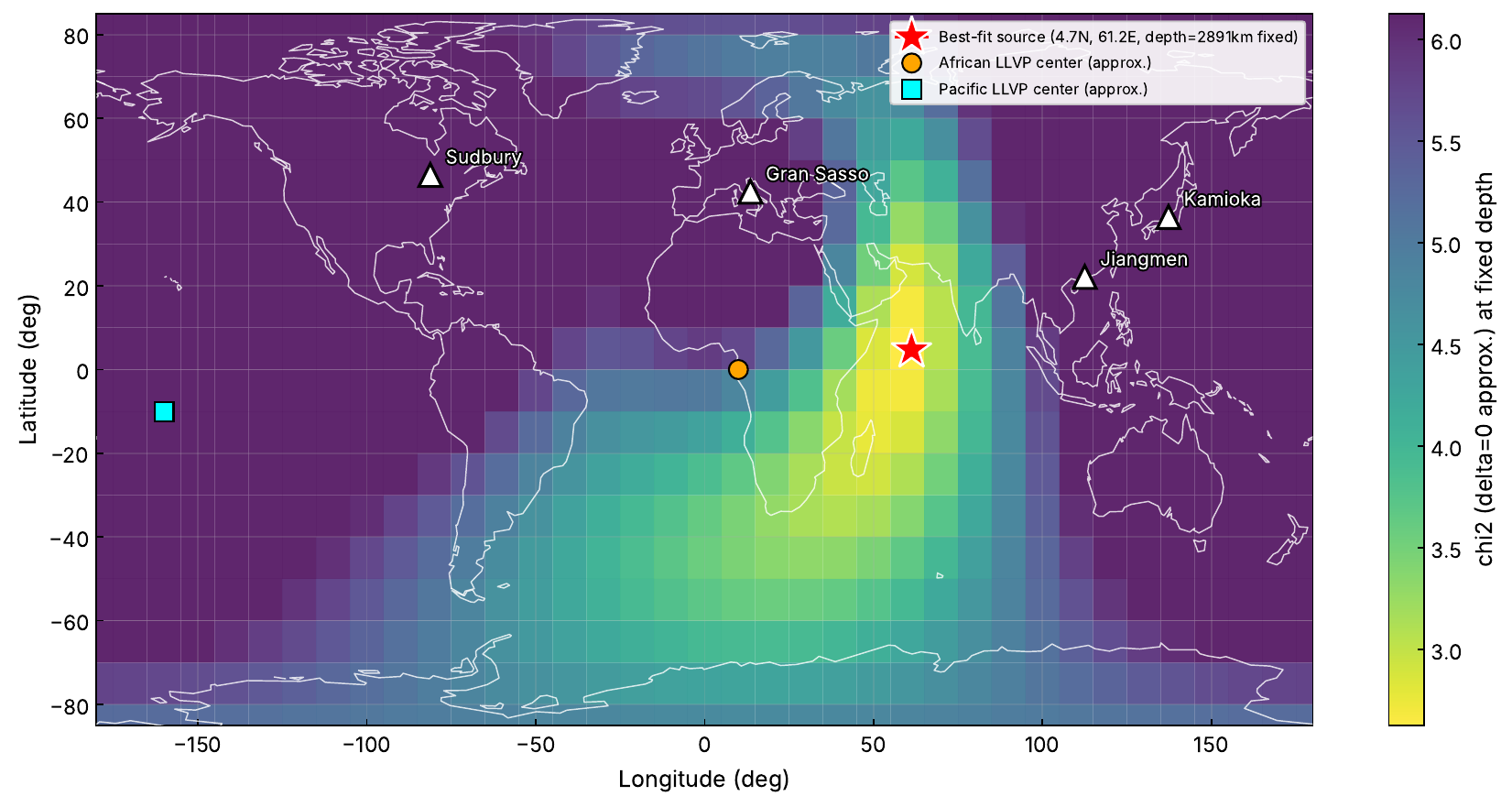}
\caption{Skymap of best fit results of point source localization at a fixed depth (2891~km) with $L$ free. The best-fit parameters are obtained at latitude = $4.72^{\circ} \pm  47.96^{\circ}$, longitude = $61.25^{\circ} \pm 21.68^{\circ}$, and $L = (5.650 \pm 4.305) \times 10^{8}$~NIU$\cdot$km$^2$, with $\chi^2 = 1.7352$.}
\label{fig:point_source_cmb}
\end{figure}

The best-fit parameters are obtained at latitude = $4.72^{\circ} \pm  47.96^{\circ}$, longitude = $61.25^{\circ} \pm 21.68^{\circ}$, and $L = (5.650 \pm 4.305) \times 10^{8}$~NIU$\cdot$km$^2$, with $\chi^2 = 1.7352$ and $p = 0.1877$ (see Figure~\ref{fig:point_source_cmb}). A robustness check yields that a global minimum exists and is repeatably found.

By fixing the source at the core–mantle boundary and letting only its position and strength float, this variant ties the depth to where structures of interest are known to exist. To test the sensitivity to this choice, we repeated the fit at fixed depths from 35 to 6300 km. Within the mantle, $\chi^2$ rises only slowly with depth, from 1.18 at 35~km to 1.74 at the CMB ($p = 0.28$ to 0.19); deeper, in the core, it rises to 3.70 at 6300 km ($p = 0.05$). The data therefore do not single out a depth within the mantle, and the CMB result is representative of mantle depths.



\section{Reference LLVP Model Fit\label{app:ref_llvp_model_fit}}

Here we report the results of our reference LLVP model fit, defined in Section~\ref{sec:llvp_models}, using the likelihood framework described in Section~\ref{sec:likelihood_framework}. Two variants are fit and reported here, each carrying the tomographic VC prior described in Section~\ref{sec:VCprior}, and the composition described in Section~\ref{sec:reservoir_composition}:
\begin{itemize}
    \item Constrained fit (geochemical Th/U term on, $w=1$): treats the bulk silicate Earth geochemistry as a genuine 1-sigma prior on BR given VC.
    \item Unconstrained fit (geochemical term off, $w=0$): removes the geochemical assumption, leaving VC and BR to be determined by the detector data and the tomographic prior alone. This result makes the fewest additional assumptions.
\end{itemize}
Here we describe the results of each of the fits and cross-check them against a full MCMC posterior. 

\subsection{Constrained Fit\label{sec:constrained_fit}}

For the constrained fit, we leave VC and BR as free parameters, coupled by the geochemical penalty at $w=1$: a genuine $1\sigma$ Gaussian prior tying BR to VC and not a hard constraint. VC is additionally pulled toward the Cottaar \& Lekic~\cite{Cottaar2016} tomographic estimate via the VC prior. This fit has two free parameters (VC, BR), as well as one lithosphere nuisance per site.

\begin{figure} [H]
\centering
\includegraphics[trim = 0mm 0mm 0mm 0mm, clip, scale=0.4]{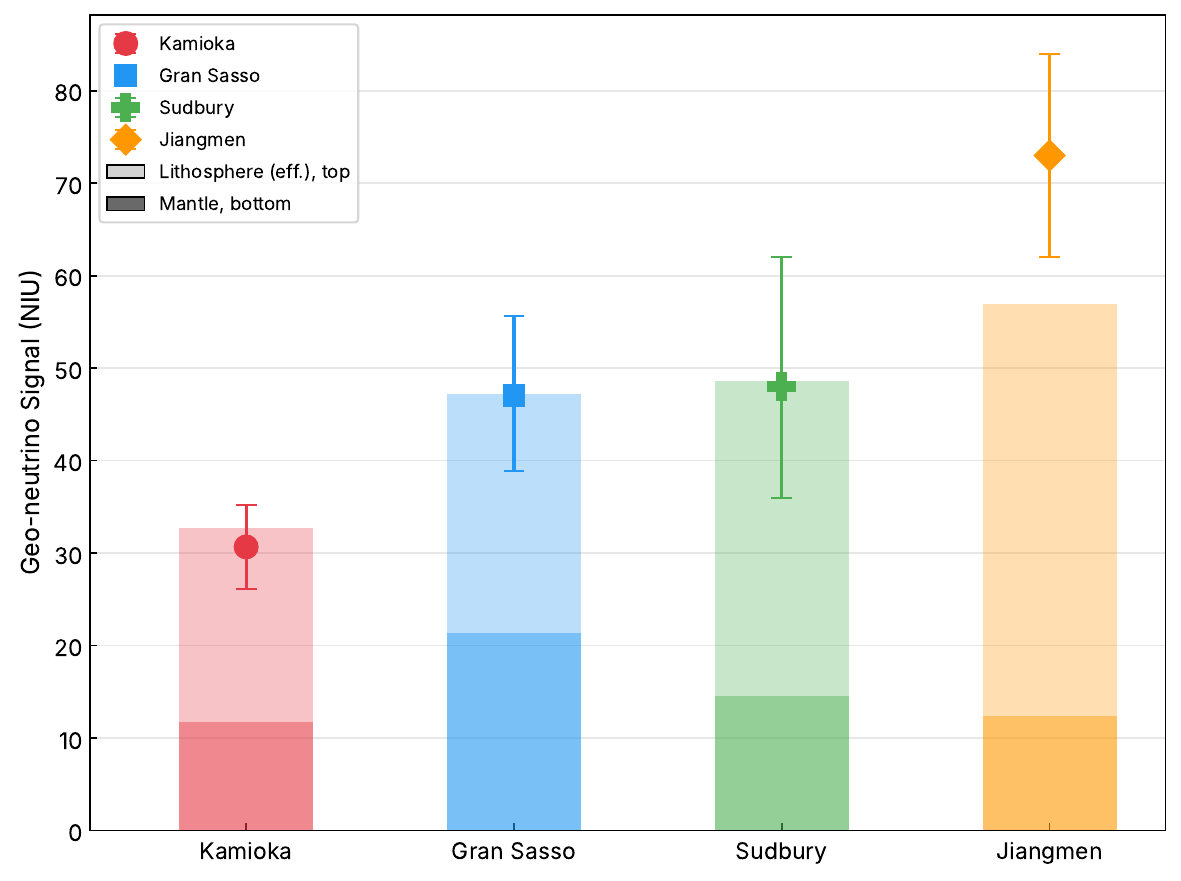}
\caption{Results of a geochemically constrained fit to our reference LLVP model, showing the predicted mantle signal at each site (dark colors), together with their effective (varied) lithosphere predictions (light colors). The best-fit values are $\text{VC} = 5.04 \pm 0.82$ and BR = ($13.9 \pm 5.3$)\%, with \chisq/\ndof\ = 3.062/4. }
\label{fig:constrained_fit_results}
\end{figure}

The results of the constrained fit are shown in Figure~\ref{fig:constrained_fit_results}. The best-fit values are $\text{VC} = 5.04 \pm 0.82$ and BR = ($13.9 \pm 5.3$)\%, with \chisq/\ndof\ = 3.06/4 and $p=0.547$. $\Delta \chi^2 = 0.721$ for one additional effective degree of freedom, compared to the uniform mantle fit. The geochemical prediction at the best-fit VC is 13.43\%, resulting in a constraint pull of $+0.070\sigma$. The lithosphere nuisance pulls are $-0.416 \sigma, -0.012 \sigma, -0.021 \sigma$, and $+0.742 \sigma$ at Kamioka, Gran Sasso, Sudbury, and Jiangmen, respectively. The fit result is a genuine global minimum: 12/12 starting points all converge to the same values.

\begin{figure} [H]
\centering
\includegraphics[trim = 0mm 0mm 0mm 0mm, clip, scale=0.44]{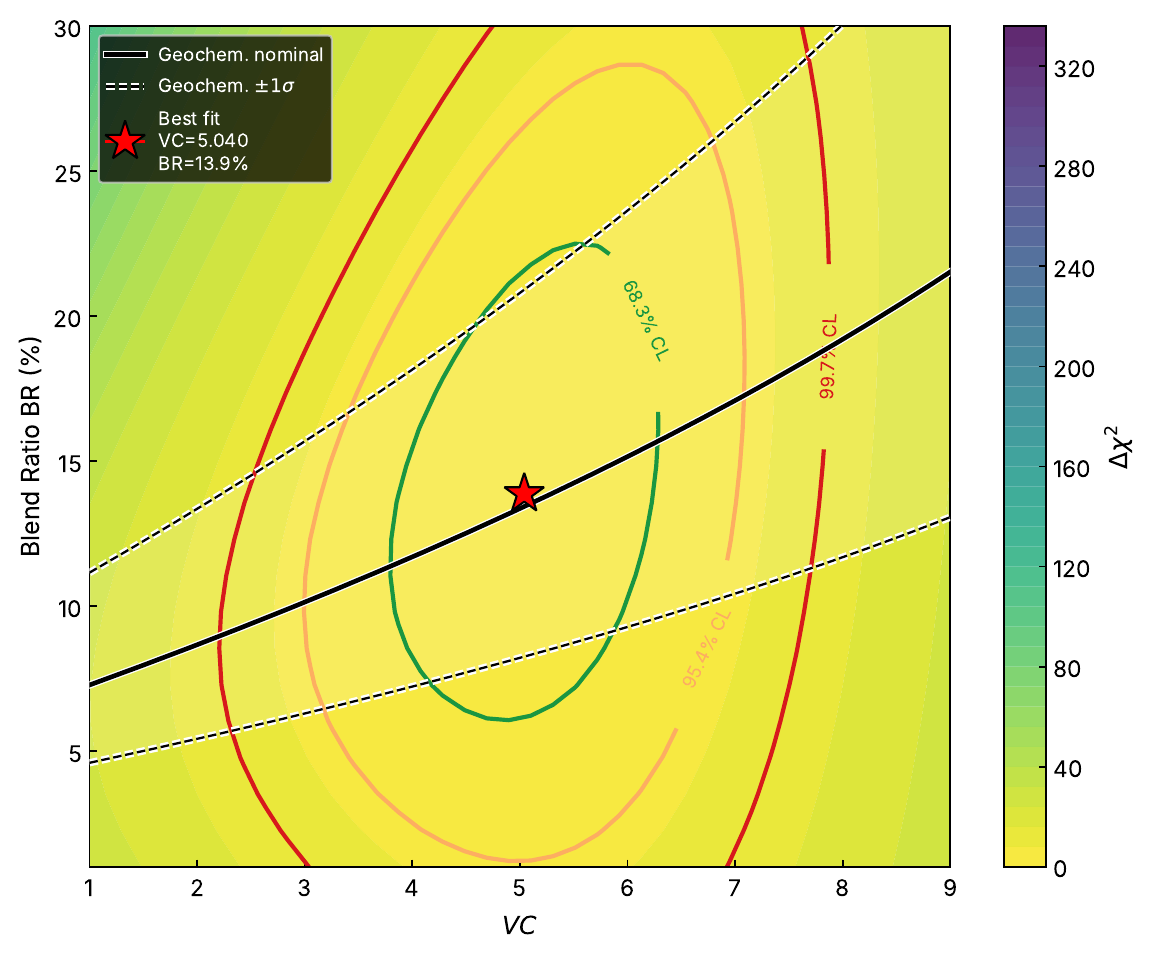}
\caption{Results of a 2D-profile-likelihood, geochemically constrained fit to our reference LLVP model.}
\label{fig:constrained_fit_results_PL}
\end{figure}

2D profile likelihood results are shown in Figure~\ref{fig:constrained_fit_results_PL}. The 2D joint minimum is in agreement with the Minuit result. Each parameter is then profiled individually: fixing one, re-optimizing everything else (including the other physical parameter and all nuisances), and then scanning the fixed parameter across its range. This gives standard 1D $\Delta \chi^2$ curves for VC and BR separately, from which $1\sigma$ and $2\sigma$ intervals can be read directly. This is the profile-likelihood analog of the Hesse errors reported by Migrad. The $1\sigma$ bounds on VC are [4.22, 5.87] and on BR [8.7, 19.4]\%. 

\subsection{Unconstrained Fit\label{sec:unconstrained_fit}}

For the unconstrained fit, VC and BR are both left as free parameters, fit by continuous 2-parameter optimization (not a grid scan), with the geochemical Th/U term switched off ($w=0$) and VC prior still on ($w_\text{VC}=1.0$). ``Unconstrained'' here means specifically not using the BSE geochemical assumption. 

Again here, the VC prior does almost all of the work of pulling VC to a well-defined value. The fitted VC uncertainty is essentially that of the prior: four combined rates with lithosphere-limited uncertainties add almost no information on VC. Although a geo-response contrast diagnostic test shows that the sites do respond differently to LLVP composition, this result is a limit of the data's precision, not an exact geometric degeneracy. The detector data alone are close to flat in VC. This result should therefore be read as ``the geoneutrino data are consistent with the independently-measured tomographic LLVP size," not as ``the geoneutrino data measure the LLVP size."
  
\begin{figure} [H]
\centering
\includegraphics[trim = 0mm 0mm 0mm 0mm, clip, scale=0.4]{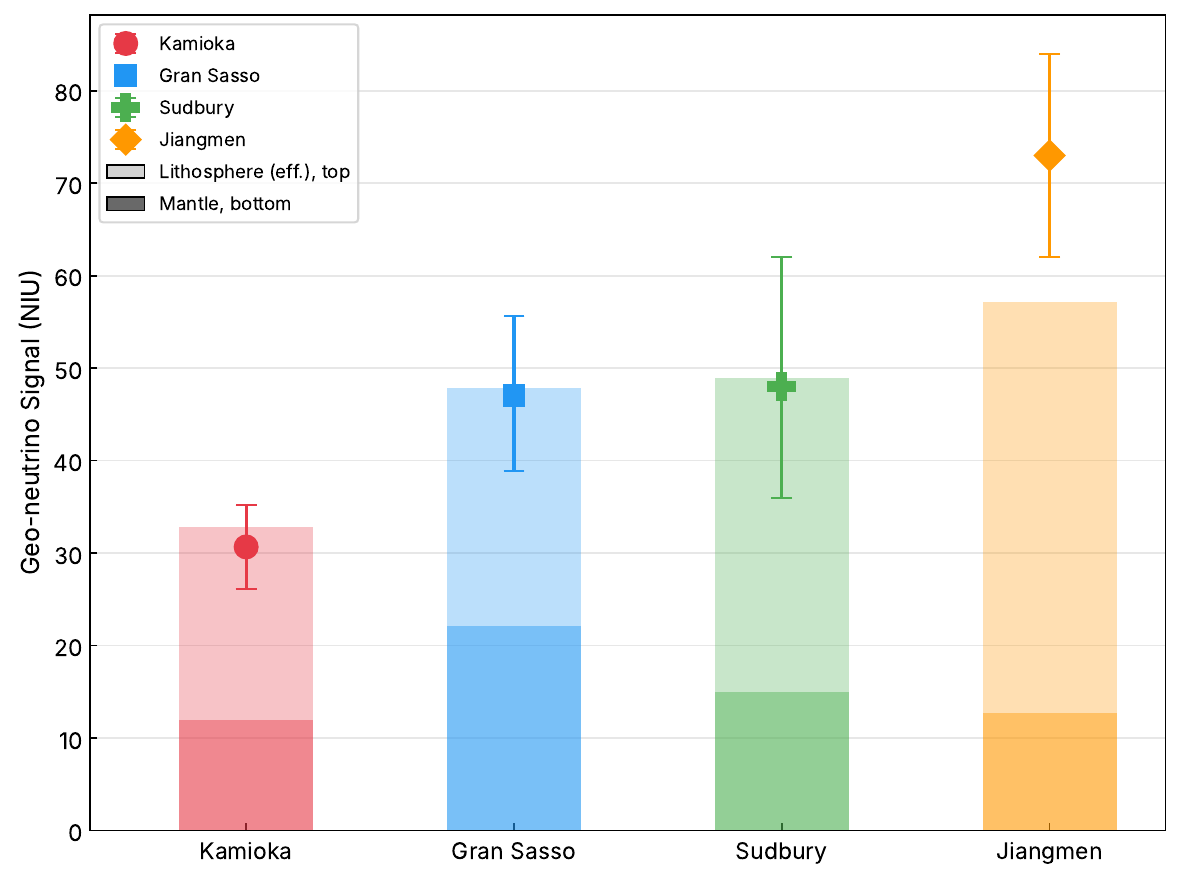}
\caption{Results of an unconstrained fit to our reference LLVP model, showing the predicted mantle signal at each site (dark colors), together with their effective (varied) lithosphere predictions (light colors). The best-fit values are VC = $5.05 \pm 0.82$ and BR = $(14.7 \pm 8.7)\%$, with \chisq/\ndof\ = 3.05/3.}
\label{fig:unconstrained_fit_results}
\end{figure}

The results of the unconstrained fit are shown in Figure~\ref{fig:unconstrained_fit_results}. The best-fit values are VC = $5.05 \pm 0.82$ and BR = $(14.7 \pm 8.7)\%$, with \chisq/\ndof\ = 3.05/3 and $p=0.384$. $\Delta \chi^2 = 0.735$ for one additional parameter, compared to the uniform mantle fit, giving a $p$-value for the improvement of 0.3901. The sigma-distance from the geochemical nominal curve is $+0.20\sigma$, compared to $+0.07\sigma$ for the constrained fit. The lithosphere nuisance pulls are $-0.442 \sigma$, $-0.046 \sigma$, $-0.029 \sigma$, and $+0.730 \sigma$ at Kamioka, Gran Sasso, Sudbury, and Jiangmen, respectively. The fit result is a genuine global minimum: 10/10 starting points all converge to the same values.

\begin{figure} [H]
\centering
\includegraphics[trim = 0mm 0mm 0mm 0mm, clip, scale=0.44]{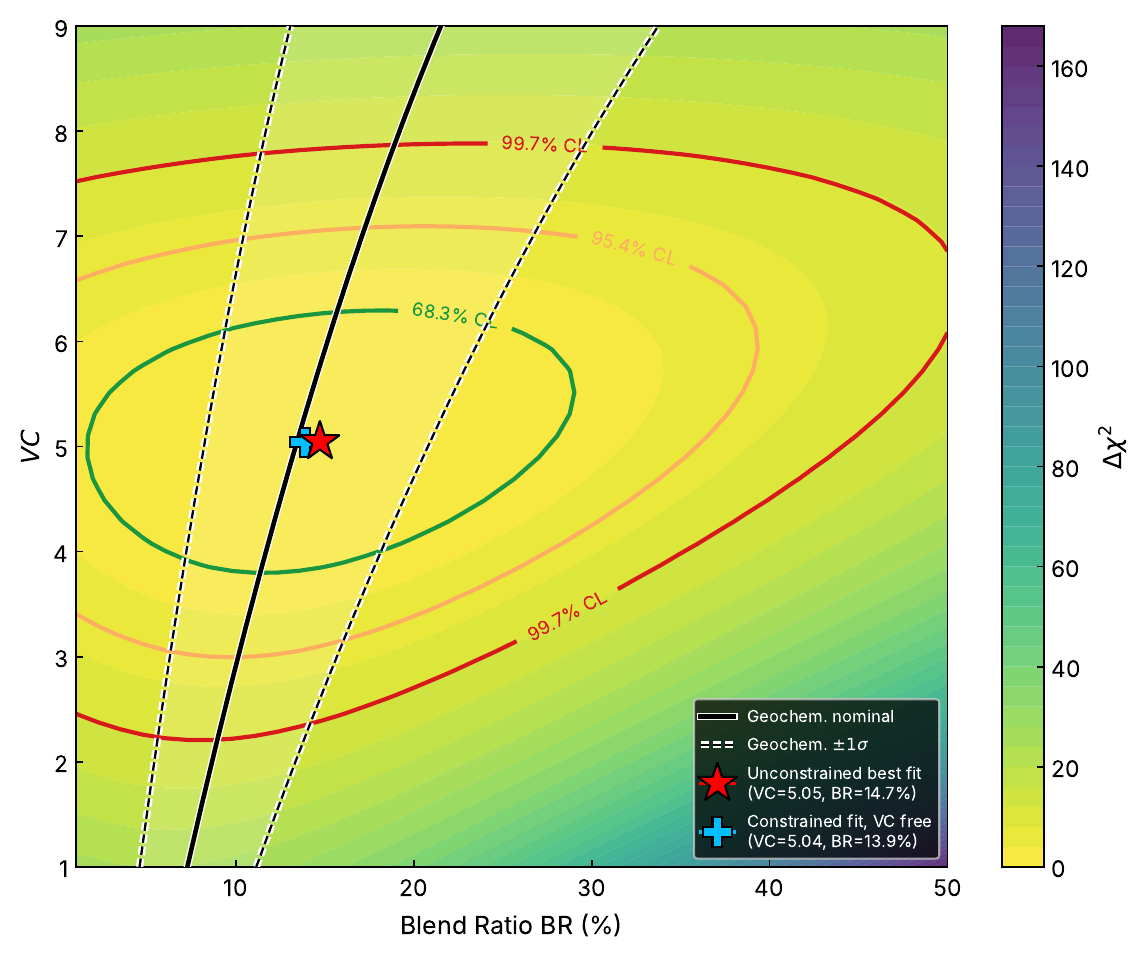}
\caption{Results of an unconstrained 2D-profile-likelihood fit to our reference LLVP model. $\Delta \chi^2$ contours for the unconstrained fit are shown, with the constrained fit's best point overlaid for direct comparison. This plot shows how much the geochemical assumption changes the result.}
\label{fig:unconstrained_fit_results_PL}
\end{figure}

2D-profile-likelihood results are shown in Figure~\ref{fig:unconstrained_fit_results_PL}, with the results of the constrained fit overlaid for comparison. The 2D joint minimum is in agreement with the Minuit result. Each parameter is then profiled individually: fixing one, re-optimizing everything else (including the other physical parameter and all nuisances), and then scanning the fixed parameter space across its range. This gives standard 1D $\Delta \chi^2$ curves for VC and BR separately, from which $1\sigma$ and $2\sigma$ intervals can be read directly. This is the profile-likelihood analog of the Hesse errors reported by Migrad. The $1\sigma$ and $2\sigma$ bounds on VC are [4.22, 5.87] and [3.39, 6.70], respectively. The $1\sigma$ and $2\sigma$ bounds on BR are [6.0, 24.0]\% and [$<1.0$, 34.1]\%, respectively. 

\subsection{MCMC Cross Check}

Migrad/Hesse gives a point estimate and a local, symmetric (parabolic) error at the minimum. As a cross-check, we sample the full posterior with \texttt{emcee}, an affine-invariant ensemble MCMC sampler, for both fits presented here, and report the asymmetric 68\% credible interval, alongside the Hesse point estimate and error. This directly tests two things Hesse cannot: whether the likelihood surface is genuinely close to parabolic near the minimum (if so, MCMC and Hesse should agree closely), and whether there is any secondary mode or boundary pile-up that Migrad's single-start local minimization could potentially miss.
                
The 68\% credible interval for VC is [4.351, 5.997] for the unconstrained fit and [4.302, 5.927] for the constrained fit. The 68\% CI for BR is [7.38, 24.45] for the unconstrained fit and [8.76, 19.51] for the constrained fit. We therefore observe close agreement between the Hesse point estimate and the MCMC median/CI. This means that the likelihood surface is close to parabolic here and confirms that Migrad found the global minimum, rather than a secondary mode. Any asymmetry in the MCMC CI beyond the symmetric Hesse error reflects genuine (mild) non-Gaussianity, not a sampling problem.

\newpage
\bibliographystyle{apsrev4-1}
\bibliography{Global_fitting}

\begin{thebibliography}{37}%
\makeatletter
\providecommand \@ifxundefined [1]{%
 \@ifx{#1\undefined}
}%
\providecommand \@ifnum [1]{%
 \ifnum #1\expandafter \@firstoftwo
 \else \expandafter \@secondoftwo
 \fi
}%
\providecommand \@ifx [1]{%
 \ifx #1\expandafter \@firstoftwo
 \else \expandafter \@secondoftwo
 \fi
}%
\providecommand \natexlab [1]{#1}%
\providecommand \enquote  [1]{``#1''}%
\providecommand \bibnamefont  [1]{#1}%
\providecommand \bibfnamefont [1]{#1}%
\providecommand \citenamefont [1]{#1}%
\providecommand \href@noop [0]{\@secondoftwo}%
\providecommand \href [0]{\begingroup \@sanitize@url \@href}%
\providecommand \@href[1]{\@@startlink{#1}\@@href}%
\providecommand \@@href[1]{\endgroup#1\@@endlink}%
\providecommand \@sanitize@url [0]{\catcode `\\12\catcode `\$12\catcode `\&12\catcode `\#12\catcode `\^12\catcode `\_12\catcode `\%12\relax}%
\providecommand \@@startlink[1]{}%
\providecommand \@@endlink[0]{}%
\providecommand \url  [0]{\begingroup\@sanitize@url \@url }%
\providecommand \@url [1]{\endgroup\@href {#1}{\urlprefix }}%
\providecommand \urlprefix  [0]{URL }%
\providecommand \Eprint [0]{\href }%
\providecommand \doibase [0]{http://dx.doi.org/}%
\providecommand \selectlanguage [0]{\@gobble}%
\providecommand \bibinfo  [0]{\@secondoftwo}%
\providecommand \bibfield  [0]{\@secondoftwo}%
\providecommand \translation [1]{[#1]}%
\providecommand \BibitemOpen [0]{}%
\providecommand \bibitemStop [0]{}%
\providecommand \bibitemNoStop [0]{.\EOS\space}%
\providecommand \EOS [0]{\spacefactor3000\relax}%
\providecommand \BibitemShut  [1]{\csname bibitem#1\endcsname}%
\let\auto@bib@innerbib\@empty
\bibitem [{\citenamefont {Abe}\ \emph {et~al.}(2022)\citenamefont {Abe} \emph {et~al.}}]{abe_etal_2022}%
  \BibitemOpen
  \bibfield  {author} {\bibinfo {author} {\bibfnamefont {S.}~\bibnamefont {Abe}} \emph {et~al.},\ }\href {\doibase 10.1029/2022GL099566} {\bibfield  {journal} {\bibinfo  {journal} {Geophys. Res. Lett.}\ }\textbf {\bibinfo {volume} {49}},\ \bibinfo {pages} {e2022GL099566} (\bibinfo {year} {2022})}\BibitemShut {NoStop}%
\bibitem [{\citenamefont {Agostini}\ \emph {et~al.}(2020)\citenamefont {Agostini} \emph {et~al.}}]{agostini_etal_2020}%
  \BibitemOpen
  \bibfield  {author} {\bibinfo {author} {\bibfnamefont {M.}~\bibnamefont {Agostini}} \emph {et~al.},\ }\href {\doibase 10.1103/PhysRevD.101.012009} {\bibfield  {journal} {\bibinfo  {journal} {Phys. Rev. D}\ }\textbf {\bibinfo {volume} {101}},\ \bibinfo {pages} {012009} (\bibinfo {year} {2020})}\BibitemShut {NoStop}%
\bibitem [{\citenamefont {Abreu}\ \emph {et~al.}(2026)\citenamefont {Abreu} \emph {et~al.}}]{SNO+:2025xxx}%
  \BibitemOpen
  \bibfield  {author} {\bibinfo {author} {\bibfnamefont {M.}~\bibnamefont {Abreu}} \emph {et~al.} (\bibinfo {collaboration} {SNO+}),\ }\href {\doibase 10.1103/mv5z-pfb7} {\bibfield  {journal} {\bibinfo  {journal} {Phys. Rev. Lett.}\ }\textbf {\bibinfo {volume} {137}},\ \bibinfo {pages} {101807} (\bibinfo {year} {2026})},\ \Eprint {http://arxiv.org/abs/2511.11856} {arXiv:2511.11856 [hep-ex]} \BibitemShut {NoStop}%
\bibitem [{\citenamefont {Wang}\ \emph {et~al.}(2026)\citenamefont {Wang} \emph {et~al.}}]{wang_etal_2026}%
  \BibitemOpen
  \bibfield  {author} {\bibinfo {author} {\bibfnamefont {Y.}~\bibnamefont {Wang}} \emph {et~al.},\ }\href@noop {} {\enquote {\bibinfo {title} {{JUNO Experiment}},}\ }\bibinfo {howpublished} {Neutrino 2026, Irvine} (\bibinfo {year} {2026})\BibitemShut {NoStop}%
\bibitem [{\citenamefont {Davies}\ and\ \citenamefont {Davies}(2010)}]{Davies2010}%
  \BibitemOpen
  \bibfield  {author} {\bibinfo {author} {\bibfnamefont {J.~H.}\ \bibnamefont {Davies}}\ and\ \bibinfo {author} {\bibfnamefont {D.~R.}\ \bibnamefont {Davies}},\ }\href {\doibase 10.5194/se-1-5-2010} {\bibfield  {journal} {\bibinfo  {journal} {Solid Earth}\ }\textbf {\bibinfo {volume} {1}},\ \bibinfo {pages} {5} (\bibinfo {year} {2010})}\BibitemShut {NoStop}%
\bibitem [{\citenamefont {Buffett}(2003)}]{buffett_2003}%
  \BibitemOpen
  \bibfield  {author} {\bibinfo {author} {\bibfnamefont {B.~A.}\ \bibnamefont {Buffett}},\ }\href {\doibase 10.1126/science.1081518} {\bibfield  {journal} {\bibinfo  {journal} {{Science}}\ }\textbf {\bibinfo {volume} {299}},\ \bibinfo {pages} {1675} (\bibinfo {year} {2003})}\BibitemShut {NoStop}%
\bibitem [{\citenamefont {{\v{S}}r{\'a}mek}\ \emph {et~al.}(2013)\citenamefont {{\v{S}}r{\'a}mek}, \citenamefont {McDonough}, \citenamefont {Kite}, \citenamefont {Lekic}, \citenamefont {Dye},\ and\ \citenamefont {Zhong}}]{sramek2013}%
  \BibitemOpen
  \bibfield  {author} {\bibinfo {author} {\bibfnamefont {O.}~\bibnamefont {{\v{S}}r{\'a}mek}}, \bibinfo {author} {\bibfnamefont {W.~F.}\ \bibnamefont {McDonough}}, \bibinfo {author} {\bibfnamefont {E.~S.}\ \bibnamefont {Kite}}, \bibinfo {author} {\bibfnamefont {V.}~\bibnamefont {Lekic}}, \bibinfo {author} {\bibfnamefont {S.~T.}\ \bibnamefont {Dye}}, \ and\ \bibinfo {author} {\bibfnamefont {S.}~\bibnamefont {Zhong}},\ }\href {\doibase 10.1016/j.epsl.2012.11.001} {\bibfield  {journal} {\bibinfo  {journal} {Earth and Planetary Science Letters}\ }\textbf {\bibinfo {volume} {361}},\ \bibinfo {pages} {356} (\bibinfo {year} {2013})}\BibitemShut {NoStop}%
\bibitem [{\citenamefont {Huang}\ \emph {et~al.}(2013)\citenamefont {Huang} \emph {et~al.}}]{huang_etal_2013}%
  \BibitemOpen
  \bibfield  {author} {\bibinfo {author} {\bibfnamefont {Y.}~\bibnamefont {Huang}} \emph {et~al.},\ }\href {\doibase doi:10.1002/ggge.20129} {\bibfield  {journal} {\bibinfo  {journal} {Geochem. Geophys. Geosyst.}\ }\textbf {\bibinfo {volume} {14}},\ \bibinfo {pages} {2003} (\bibinfo {year} {2013})}\BibitemShut {NoStop}%
\bibitem [{Note1()}]{Note1}%
  \BibitemOpen
  \bibinfo {note} {1 NIU is equal to 1 IBD interaction per year exposure of $10^{32}$ free protons and is equivalent to $1$~Terrestrial Neutrino Unit (TNU)}\BibitemShut {NoStop}%
\bibitem [{\citenamefont {Gao}\ \emph {et~al.}(2020)\citenamefont {Gao}, \citenamefont {Li}, \citenamefont {Han}, \citenamefont {Wang}, \citenamefont {Li}, \citenamefont {Xi}, \citenamefont {Liu}, \citenamefont {Mao}, \citenamefont {Sun},\ and\ \citenamefont {Xu}}]{Gao2020JULOC}%
  \BibitemOpen
  \bibfield  {author} {\bibinfo {author} {\bibfnamefont {R.}~\bibnamefont {Gao}}, \bibinfo {author} {\bibfnamefont {Z.}~\bibnamefont {Li}}, \bibinfo {author} {\bibfnamefont {R.}~\bibnamefont {Han}}, \bibinfo {author} {\bibfnamefont {A.}~\bibnamefont {Wang}}, \bibinfo {author} {\bibfnamefont {Y.-F.}\ \bibnamefont {Li}}, \bibinfo {author} {\bibfnamefont {Y.}~\bibnamefont {Xi}}, \bibinfo {author} {\bibfnamefont {J.}~\bibnamefont {Liu}}, \bibinfo {author} {\bibfnamefont {X.}~\bibnamefont {Mao}}, \bibinfo {author} {\bibfnamefont {Y.}~\bibnamefont {Sun}}, \ and\ \bibinfo {author} {\bibfnamefont {Y.}~\bibnamefont {Xu}},\ }\href {\doibase 10.1016/j.pepi.2019.106409} {\bibfield  {journal} {\bibinfo  {journal} {Physics of the Earth and Planetary Interiors}\ }\textbf {\bibinfo {volume} {299}},\ \bibinfo {pages} {106409} (\bibinfo {year} {2020})}\BibitemShut {NoStop}%
\bibitem [{\citenamefont {Strati}\ \emph {et~al.}(2017)\citenamefont {Strati} \emph {et~al.}}]{strati_etal_2017}%
  \BibitemOpen
  \bibfield  {author} {\bibinfo {author} {\bibfnamefont {V.}~\bibnamefont {Strati}} \emph {et~al.},\ }\href {\doibase https://doi.org/10.1002/2017GC007067} {\bibfield  {journal} {\bibinfo  {journal} {Geochem. Geophys. Geosyst.}\ }\textbf {\bibinfo {volume} {18}},\ \bibinfo {pages} {4326} (\bibinfo {year} {2017})}\BibitemShut {NoStop}%
\bibitem [{\citenamefont {Laursen}\ \emph {et~al.}(2019)\citenamefont {Laursen} \emph {et~al.}}]{laursen_etal_2019}%
  \BibitemOpen
  \bibfield  {author} {\bibinfo {author} {\bibfnamefont {P.}~\bibnamefont {Laursen}} \emph {et~al.},\ }\href {\doibase 10.1051/0004-6361/201833645} {\bibfield  {journal} {\bibinfo  {journal} {Astronomy and Astrophysics}\ }\textbf {\bibinfo {volume} {627}},\ \bibinfo {pages} {A84} (\bibinfo {year} {2019})}\BibitemShut {NoStop}%
\bibitem [{\citenamefont {van~der Meer}\ \emph {et~al.}(2018)\citenamefont {van~der Meer}, \citenamefont {van Hinsbergen},\ and\ \citenamefont {Spakman}}]{vanderMeer2018}%
  \BibitemOpen
  \bibfield  {author} {\bibinfo {author} {\bibfnamefont {D.~G.}\ \bibnamefont {van~der Meer}}, \bibinfo {author} {\bibfnamefont {D.~J.~J.}\ \bibnamefont {van Hinsbergen}}, \ and\ \bibinfo {author} {\bibfnamefont {W.}~\bibnamefont {Spakman}},\ }\href {\doibase 10.1016/j.tecto.2017.10.004} {\bibfield  {journal} {\bibinfo  {journal} {Tectonophysics}\ }\textbf {\bibinfo {volume} {723}},\ \bibinfo {pages} {309} (\bibinfo {year} {2018})}\BibitemShut {NoStop}%
\bibitem [{\citenamefont {Plank}\ and\ \citenamefont {Langmuir}(1998)}]{Plank1998}%
  \BibitemOpen
  \bibfield  {author} {\bibinfo {author} {\bibfnamefont {T.}~\bibnamefont {Plank}}\ and\ \bibinfo {author} {\bibfnamefont {C.~H.}\ \bibnamefont {Langmuir}},\ }\href {\doibase 10.1016/S0009-2541(97)00150-2} {\bibfield  {journal} {\bibinfo  {journal} {Chemical Geology}\ }\textbf {\bibinfo {volume} {145}},\ \bibinfo {pages} {325} (\bibinfo {year} {1998})}\BibitemShut {NoStop}%
\bibitem [{\citenamefont {Workman}\ and\ \citenamefont {Hart}(2005)}]{workman_hart_2005}%
  \BibitemOpen
  \bibfield  {author} {\bibinfo {author} {\bibfnamefont {R.~K.}\ \bibnamefont {Workman}}\ and\ \bibinfo {author} {\bibfnamefont {S.~R.}\ \bibnamefont {Hart}},\ }\href {\doibase doi:10.1016/j.epsl.2004.12.005} {\bibfield  {journal} {\bibinfo  {journal} {Earth Planet. Sci. Lett.}\ }\textbf {\bibinfo {volume} {231}},\ \bibinfo {pages} {53 } (\bibinfo {year} {2005})}\BibitemShut {NoStop}%
\bibitem [{\citenamefont {Salters}\ and\ \citenamefont {Stracke}(2004)}]{salters_stracke_2004}%
  \BibitemOpen
  \bibfield  {author} {\bibinfo {author} {\bibfnamefont {V.~J.~M.}\ \bibnamefont {Salters}}\ and\ \bibinfo {author} {\bibfnamefont {A.}~\bibnamefont {Stracke}},\ }\href {\doibase https://doi.org/10.1029/2003GC000597} {\bibfield  {journal} {\bibinfo  {journal} {Geochem. Geophys. Geosyst.}\ }\textbf {\bibinfo {volume} {5}},\ \bibinfo {pages} {Q05B07} (\bibinfo {year} {2004})}\BibitemShut {NoStop}%
\bibitem [{\citenamefont {Arevalo}\ and\ \citenamefont {McDonough}(2010)}]{Arevalo2010Chemical}%
  \BibitemOpen
  \bibfield  {author} {\bibinfo {author} {\bibfnamefont {R.}~\bibnamefont {Arevalo}}\ and\ \bibinfo {author} {\bibfnamefont {W.~F.}\ \bibnamefont {McDonough}},\ }\href {\doibase 10.1016/j.chemgeo.2009.12.010} {\bibfield  {journal} {\bibinfo  {journal} {Chemical Geology}\ }\textbf {\bibinfo {volume} {271}},\ \bibinfo {pages} {70} (\bibinfo {year} {2010})}\BibitemShut {NoStop}%
\bibitem [{\citenamefont {Dziewonski}\ and\ \citenamefont {Anderson}(1981)}]{dziew81}%
  \BibitemOpen
  \bibfield  {author} {\bibinfo {author} {\bibfnamefont {A.~M.}\ \bibnamefont {Dziewonski}}\ and\ \bibinfo {author} {\bibfnamefont {D.~L.}\ \bibnamefont {Anderson}},\ }\href {\doibase 10.1016/0031-9201(81)90046-7} {\bibfield  {journal} {\bibinfo  {journal} {Phys. Earth Planet. Inter.}\ }\textbf {\bibinfo {volume} {25}},\ \bibinfo {pages} {297} (\bibinfo {year} {1981})}\BibitemShut {NoStop}%
\bibitem [{\citenamefont {Wipperfurth}\ \emph {et~al.}(2018)\citenamefont {Wipperfurth} \emph {et~al.}}]{wipperfurth_etal_2018}%
  \BibitemOpen
  \bibfield  {author} {\bibinfo {author} {\bibfnamefont {S.~A.}\ \bibnamefont {Wipperfurth}} \emph {et~al.},\ }\href {\doibase https://doi.org/10.1016/j.epsl.2018.06.029} {\bibfield  {journal} {\bibinfo  {journal} {Earth Planet. Sci. Lett.}\ }\textbf {\bibinfo {volume} {498}},\ \bibinfo {pages} {196 } (\bibinfo {year} {2018})}\BibitemShut {NoStop}%
\bibitem [{\citenamefont {Yuan}\ and\ \citenamefont {Li}(2022)}]{yuan_li_2022}%
  \BibitemOpen
  \bibfield  {author} {\bibinfo {author} {\bibfnamefont {Q.}~\bibnamefont {Yuan}}\ and\ \bibinfo {author} {\bibfnamefont {M.}~\bibnamefont {Li}},\ }\href {\doibase https://doi.org/10.1038/s41561-022-00908-3} {\bibfield  {journal} {\bibinfo  {journal} {Nat. Geosci.}\ }\textbf {\bibinfo {volume} {15}},\ \bibinfo {pages} {334 } (\bibinfo {year} {2022})}\BibitemShut {NoStop}%
\bibitem [{\citenamefont {He}\ and\ \citenamefont {Wen}(2009)}]{he_wen_2009}%
  \BibitemOpen
  \bibfield  {author} {\bibinfo {author} {\bibfnamefont {Y.}~\bibnamefont {He}}\ and\ \bibinfo {author} {\bibfnamefont {L.}~\bibnamefont {Wen}},\ }\href {\doibase doi:10.1029/2008JB005814} {\bibfield  {journal} {\bibinfo  {journal} {J. Geophys. Res.}\ }\textbf {\bibinfo {volume} {114}},\ \bibinfo {pages} {B02309} (\bibinfo {year} {2009})}\BibitemShut {NoStop}%
\bibitem [{\citenamefont {Wang}\ and\ \citenamefont {Wen}(2007)}]{wang_wen_2007}%
  \BibitemOpen
  \bibfield  {author} {\bibinfo {author} {\bibfnamefont {Y.}~\bibnamefont {Wang}}\ and\ \bibinfo {author} {\bibfnamefont {L.}~\bibnamefont {Wen}},\ }\href {\doibase doi:10.1029/2006JB004483} {\bibfield  {journal} {\bibinfo  {journal} {J. Geophys. Res.}\ }\textbf {\bibinfo {volume} {112}},\ \bibinfo {pages} {B05313} (\bibinfo {year} {2007})}\BibitemShut {NoStop}%
\bibitem [{\citenamefont {Hosseini}\ \emph {et~al.}(2018)\citenamefont {Hosseini}, \citenamefont {Matthews}, \citenamefont {Sigloch}, \citenamefont {Shephard}, \citenamefont {Domeier},\ and\ \citenamefont {Tsekhmistrenko}}]{SubMachine}%
  \BibitemOpen
  \bibfield  {author} {\bibinfo {author} {\bibfnamefont {K.}~\bibnamefont {Hosseini}}, \bibinfo {author} {\bibfnamefont {K.~J.}\ \bibnamefont {Matthews}}, \bibinfo {author} {\bibfnamefont {K.}~\bibnamefont {Sigloch}}, \bibinfo {author} {\bibfnamefont {G.~E.}\ \bibnamefont {Shephard}}, \bibinfo {author} {\bibfnamefont {M.}~\bibnamefont {Domeier}}, \ and\ \bibinfo {author} {\bibfnamefont {M.}~\bibnamefont {Tsekhmistrenko}},\ }\href {\doibase 10.1029/2018GC007431} {\bibfield  {journal} {\bibinfo  {journal} {Geochemistry, Geophysics, Geosystems}\ }\textbf {\bibinfo {volume} {19}},\ \bibinfo {pages} {1464} (\bibinfo {year} {2018})}\BibitemShut {NoStop}%
\bibitem [{\citenamefont {Shephard}\ \emph {et~al.}(2017)\citenamefont {Shephard}, \citenamefont {Matthews}, \citenamefont {Hosseini},\ and\ \citenamefont {Domeier}}]{Shephard_2017}%
  \BibitemOpen
  \bibfield  {author} {\bibinfo {author} {\bibfnamefont {G.~E.}\ \bibnamefont {Shephard}}, \bibinfo {author} {\bibfnamefont {K.~J.}\ \bibnamefont {Matthews}}, \bibinfo {author} {\bibfnamefont {K.}~\bibnamefont {Hosseini}}, \ and\ \bibinfo {author} {\bibfnamefont {M.}~\bibnamefont {Domeier}},\ }\href {\doibase 10.1038/s41598-017-11039-w} {\bibfield  {journal} {\bibinfo  {journal} {Scientific Reports}\ }\textbf {\bibinfo {volume} {7}},\ \bibinfo {pages} {10976} (\bibinfo {year} {2017})}\BibitemShut {NoStop}%
\bibitem [{\citenamefont {Cottaar}\ and\ \citenamefont {Lekic}(2016)}]{Cottaar2016}%
  \BibitemOpen
  \bibfield  {author} {\bibinfo {author} {\bibfnamefont {S.}~\bibnamefont {Cottaar}}\ and\ \bibinfo {author} {\bibfnamefont {V.}~\bibnamefont {Lekic}},\ }\href {\doibase 10.1093/gji/ggw324} {\bibfield  {journal} {\bibinfo  {journal} {Geophysical Journal International}\ }\textbf {\bibinfo {volume} {207}},\ \bibinfo {pages} {1122} (\bibinfo {year} {2016})},\ \Eprint {http://arxiv.org/abs/https://academic.oup.com/gji/article-pdf/207/2/1122/28502525/ggw324.pdf} {https://academic.oup.com/gji/article-pdf/207/2/1122/28502525/ggw324.pdf} \BibitemShut {NoStop}%
\bibitem [{\citenamefont {Panton}\ \emph {et~al.}(2025)\citenamefont {Panton} \emph {et~al.}}]{panton_etal}%
  \BibitemOpen
  \bibfield  {author} {\bibinfo {author} {\bibfnamefont {J.}~\bibnamefont {Panton}} \emph {et~al.},\ }\href {\doibase https://doi.org/10.1038/s41598-025-88931-3} {\bibfield  {journal} {\bibinfo  {journal} {Nat. Sci. Rep.}\ }\textbf {\bibinfo {volume} {15}},\ \bibinfo {pages} {4456} (\bibinfo {year} {2025})}\BibitemShut {NoStop}%
\bibitem [{\citenamefont {Krymarys}\ \emph {et~al.}(2025)\citenamefont {Krymarys}, \citenamefont {Murakami}, \citenamefont {Saha},\ and\ \citenamefont {Liebske}}]{Krymarys2025}%
  \BibitemOpen
  \bibfield  {author} {\bibinfo {author} {\bibfnamefont {E.}~\bibnamefont {Krymarys}}, \bibinfo {author} {\bibfnamefont {M.}~\bibnamefont {Murakami}}, \bibinfo {author} {\bibfnamefont {P.}~\bibnamefont {Saha}}, \ and\ \bibinfo {author} {\bibfnamefont {C.}~\bibnamefont {Liebske}},\ }\href {\doibase 10.1038/s43247-025-02700-8} {\bibfield  {journal} {\bibinfo  {journal} {Communications Earth \& Environment}\ }\textbf {\bibinfo {volume} {6}},\ \bibinfo {pages} {713} (\bibinfo {year} {2025})}\BibitemShut {NoStop}%
\bibitem [{\citenamefont {Jackson}\ \emph {et~al.}(2021)\citenamefont {Jackson} \emph {et~al.}}]{jackson_etal_2021}%
  \BibitemOpen
  \bibfield  {author} {\bibinfo {author} {\bibfnamefont {M.~G.}\ \bibnamefont {Jackson}} \emph {et~al.},\ }\href {\doibase https://doi.org/10.1029/2020GC009525} {\bibfield  {journal} {\bibinfo  {journal} {Geochem. Geophys. Geosyst.}\ }\textbf {\bibinfo {volume} {22}},\ \bibinfo {pages} {e2020GC009525} (\bibinfo {year} {2021})}\BibitemShut {NoStop}%
\bibitem [{\citenamefont {Doucet}\ \emph {et~al.}(2020)\citenamefont {Doucet}, \citenamefont {Li}, \citenamefont {Gamal El~Dien} \emph {et~al.}}]{doucet2020}%
  \BibitemOpen
  \bibfield  {author} {\bibinfo {author} {\bibfnamefont {L.-S.}\ \bibnamefont {Doucet}}, \bibinfo {author} {\bibfnamefont {Z.-X.}\ \bibnamefont {Li}}, \bibinfo {author} {\bibfnamefont {H.}~\bibnamefont {Gamal El~Dien}},  \emph {et~al.},\ }\href {\doibase 10.1038/s41561-020-0599-9} {\bibfield  {journal} {\bibinfo  {journal} {Nature Geoscience}\ }\textbf {\bibinfo {volume} {13}},\ \bibinfo {pages} {611} (\bibinfo {year} {2020})}\BibitemShut {NoStop}%
\bibitem [{\citenamefont {Hirose}\ \emph {et~al.}(2005)\citenamefont {Hirose} \emph {et~al.}}]{hirose_etal_2005}%
  \BibitemOpen
  \bibfield  {author} {\bibinfo {author} {\bibfnamefont {K.}~\bibnamefont {Hirose}} \emph {et~al.},\ }\href {\doibase https://doi.org/10.1016/j.epsl.2005.06.035} {\bibfield  {journal} {\bibinfo  {journal} {Earth Plant. Sci. Lett.}\ }\textbf {\bibinfo {volume} {237}},\ \bibinfo {pages} {239 } (\bibinfo {year} {2005})}\BibitemShut {NoStop}%
\bibitem [{\citenamefont {Li}\ and\ \citenamefont {McNamara}(2013)}]{li_mcnamara_2013}%
  \BibitemOpen
  \bibfield  {author} {\bibinfo {author} {\bibfnamefont {M.}~\bibnamefont {Li}}\ and\ \bibinfo {author} {\bibfnamefont {A.~K.}\ \bibnamefont {McNamara}},\ }\href {\doibase 10.1002/jgrb.50156} {\bibfield  {journal} {\bibinfo  {journal} {J. Geophys. Res. Solid Earth}\ }\textbf {\bibinfo {volume} {118}},\ \bibinfo {pages} {1807 } (\bibinfo {year} {2013})}\BibitemShut {NoStop}%
\bibitem [{\citenamefont {Pasyanos}\ \emph {et~al.}(2014)\citenamefont {Pasyanos}, \citenamefont {Masters}, \citenamefont {Laske},\ and\ \citenamefont {Ma}}]{Pasyanos2014LITHO10}%
  \BibitemOpen
  \bibfield  {author} {\bibinfo {author} {\bibfnamefont {M.~E.}\ \bibnamefont {Pasyanos}}, \bibinfo {author} {\bibfnamefont {T.~G.}\ \bibnamefont {Masters}}, \bibinfo {author} {\bibfnamefont {G.}~\bibnamefont {Laske}}, \ and\ \bibinfo {author} {\bibfnamefont {Z.}~\bibnamefont {Ma}},\ }\href {\doibase 10.1002/2013JB010626} {\bibfield  {journal} {\bibinfo  {journal} {Journal of Geophysical Research: Solid Earth}\ }\textbf {\bibinfo {volume} {119}},\ \bibinfo {pages} {2153} (\bibinfo {year} {2014})}\BibitemShut {NoStop}%
\bibitem [{\citenamefont {Coltorti}\ \emph {et~al.}(2011)\citenamefont {Coltorti}, \citenamefont {Boraso}, \citenamefont {Mantovani}, \citenamefont {Morsilli}, \citenamefont {Fiorentini}, \citenamefont {Riva}, \citenamefont {Rusciadelli}, \citenamefont {Tassinari}, \citenamefont {Tomei}, \citenamefont {Di~Carlo},\ and\ \citenamefont {Chubakov}}]{Coltorti_2011}%
  \BibitemOpen
  \bibfield  {author} {\bibinfo {author} {\bibfnamefont {M.}~\bibnamefont {Coltorti}}, \bibinfo {author} {\bibfnamefont {R.}~\bibnamefont {Boraso}}, \bibinfo {author} {\bibfnamefont {F.}~\bibnamefont {Mantovani}}, \bibinfo {author} {\bibfnamefont {M.}~\bibnamefont {Morsilli}}, \bibinfo {author} {\bibfnamefont {G.}~\bibnamefont {Fiorentini}}, \bibinfo {author} {\bibfnamefont {A.}~\bibnamefont {Riva}}, \bibinfo {author} {\bibfnamefont {G.}~\bibnamefont {Rusciadelli}}, \bibinfo {author} {\bibfnamefont {R.}~\bibnamefont {Tassinari}}, \bibinfo {author} {\bibfnamefont {C.}~\bibnamefont {Tomei}}, \bibinfo {author} {\bibfnamefont {G.}~\bibnamefont {Di~Carlo}}, \ and\ \bibinfo {author} {\bibfnamefont {V.}~\bibnamefont {Chubakov}},\ }\href {\doibase 10.1016/j.gca.2011.01.024} {\bibfield  {journal} {\bibinfo  {journal} {Geochimica et Cosmochimica Acta}\ }\textbf {\bibinfo {volume} {75}},\ \bibinfo {pages} {2271–2294} (\bibinfo {year} {2011})}\BibitemShut {NoStop}%
\bibitem [{\citenamefont {Enomoto}\ \emph {et~al.}(2007)\citenamefont {Enomoto}, \citenamefont {Ohtani}, \citenamefont {Inoue},\ and\ \citenamefont {Suzuki}}]{ENOMOTO2007147}%
  \BibitemOpen
  \bibfield  {author} {\bibinfo {author} {\bibfnamefont {S.}~\bibnamefont {Enomoto}}, \bibinfo {author} {\bibfnamefont {E.}~\bibnamefont {Ohtani}}, \bibinfo {author} {\bibfnamefont {K.}~\bibnamefont {Inoue}}, \ and\ \bibinfo {author} {\bibfnamefont {A.}~\bibnamefont {Suzuki}},\ }\href {\doibase https://doi.org/10.1016/j.epsl.2007.03.038} {\bibfield  {journal} {\bibinfo  {journal} {Earth and Planetary Science Letters}\ }\textbf {\bibinfo {volume} {258}},\ \bibinfo {pages} {147} (\bibinfo {year} {2007})}\BibitemShut {NoStop}%
\bibitem [{\citenamefont {Gale}\ \emph {et~al.}(2013)\citenamefont {Gale}, \citenamefont {Dalton}, \citenamefont {Langmuir}, \citenamefont {Su},\ and\ \citenamefont {Schilling}}]{Gale2013}%
  \BibitemOpen
  \bibfield  {author} {\bibinfo {author} {\bibfnamefont {A.}~\bibnamefont {Gale}}, \bibinfo {author} {\bibfnamefont {C.~A.}\ \bibnamefont {Dalton}}, \bibinfo {author} {\bibfnamefont {C.~H.}\ \bibnamefont {Langmuir}}, \bibinfo {author} {\bibfnamefont {Y.}~\bibnamefont {Su}}, \ and\ \bibinfo {author} {\bibfnamefont {J.-G.}\ \bibnamefont {Schilling}},\ }\href {\doibase 10.1029/2012gc004334} {\bibfield  {journal} {\bibinfo  {journal} {Geochemistry, Geophysics, Geosystems}\ }\textbf {\bibinfo {volume} {14}},\ \bibinfo {pages} {489} (\bibinfo {year} {2013})}\BibitemShut {NoStop}%
\bibitem [{\citenamefont {Scholl}\ and\ \citenamefont {von Huene}(2009)}]{Scholl2009}%
  \BibitemOpen
  \bibfield  {author} {\bibinfo {author} {\bibfnamefont {D.~W.}\ \bibnamefont {Scholl}}\ and\ \bibinfo {author} {\bibfnamefont {R.}~\bibnamefont {von Huene}},\ }in\ \href {\doibase 10.1144/SP318.4} {\emph {\bibinfo {booktitle} {Earth Accretionary Systems in Space and Time}}},\ \bibinfo {series} {Geological Society Special Publication}, Vol.\ \bibinfo {volume} {318},\ \bibinfo {editor} {edited by\ \bibinfo {editor} {\bibfnamefont {P.~A.}\ \bibnamefont {Cawood}}\ and\ \bibinfo {editor} {\bibfnamefont {A.}~\bibnamefont {Kr{\"o}ner}}}\ (\bibinfo  {publisher} {Geological Society of London},\ \bibinfo {year} {2009})\ pp.\ \bibinfo {pages} {105--125}\BibitemShut {NoStop}%
\bibitem [{\citenamefont {Stern}(2011)}]{Stern2011SubductionER}%
  \BibitemOpen
  \bibfield  {author} {\bibinfo {author} {\bibfnamefont {C.~R.}\ \bibnamefont {Stern}},\ }\href {\doibase 10.1016/j.gr.2011.03.006} {\bibfield  {journal} {\bibinfo  {journal} {Gondwana Research}\ }\textbf {\bibinfo {volume} {20}},\ \bibinfo {pages} {284} (\bibinfo {year} {2011})}\BibitemShut {NoStop}%
\end{thebibliography}%

\end{document}